\documentclass[oldversion]{aa}
\usepackage{graphicx}
\usepackage{psfig}
\begin{document}

\title{The nature and evolution of Nova Cygni 2006\thanks{Tables~1, 3 and Figures 7, 8, 16-20 available 
       only electronically}$^{,}$\thanks{Based in part on observations
       obtained with the Asiago telescopes}}

   \author{U.~Munari\inst{1}, A.~Siviero\inst{1}, A.~Henden\inst{2}, 
           G.~Cardarelli\inst{3}, 
           G.~Cherini\inst{3}, 
           S.~Dallaporta\inst{3}, 
           G.~Dalla Via\inst{3}, 
           A.~Frigo\inst{3},
           R.~Jurdana-Sepi\v{c},
           S.~Moretti\inst{3},
           P.~Ochner\inst{3}, 
           S.~Tomaselli\inst{3}, 
           S.~Tomasoni\inst{3}, 
           P.~Valisa\inst{3}, H.~Navasardyan\inst{1}, M.~Valentini\inst{4,1}}

   \institute{INAF Osservatorio Astronomico di Padova, via dell'Osservatorio 8, 36012 Asiago (VI), Italy
	      \and
              AAVSO, 49 Bay State Road, Cambridge, MA 02138, USA
              \and
              ANS Collaboration, c/o Osservatorio Astronomico, via dell'Osservatorio 8, 36012 Asiago (VI), Italy
              \and
              Isaac Newton Group of Telescopes, Apartado de Correos 321, E-38700 Santa Cruz de La Palma, Spain
              }

   \date{Received Feb 2, 2008; accepted YYY ZZ, XXXX}

     \abstract{
  {\it Aims}. Nova Cyg 2006 has been intensively observed throughout its
              full outburst. We investigate the energetics and evolution of the
              central source and of the expanding ejecta, their chemical
              abundances and ionization structure, and the formation of dust.\\
 {\it Methods}. We recorded low, medium, and/or high-resolution spectra
              (calibrated into accurate absolute fluxes) on 39 nights,
              along with 2353 photometric UBVRcIc measures on 313 nights, and
              complemented them with IR data from the literature.\\
  {\it Results}. The nova displayed initially the normal photometric and
              spectroscopic evolution of a fast nova of the FeII-type. 
              Pre-maximum, principal, diffuse-enhanced, and Orion absorption
              systems developed in a normal way. After the initial outburst,
              the nova progressively slowed its fading pace until the
              decline reversed and a second maximum was reached (eight
              months later), accompanied by large spectroscopic changes.
              Following the rapid decline from second maximum, the nova
              finally entered the nebular phase and formed optically thin
              dust. We performed a photo-ionization analysis of the
              emission-line spectrum during the nebular phase, which showed a
              strong enrichment of the ejecta in nitrogen and oxygen, and
              none in neon, in agreement with theoretical predictions for
              the estimated 1.0~M$_\odot$ white dwarf in Nova Cyg 2006.  The
              similarities with the poorly investigated V1493 Nova Aql 1999a
              are discussed.}

    \keywords{Stars: novae -- Stars: individual: V2362 Cyg
              }

	\authorrunning{U. Munari et al.}
	\titlerunning{Nova Cyg 2006}

   \maketitle

\section{Introduction}

Classical novae are powered by thermonuclear runaways occurring on white
dwarfs in close binary systems of the cataclysmic variable type, during
which H-rich material accreted from a low-mass companion is burned and
explosively ejected into the interstellar medium, typically at $\sim$1000
km~sec$^{-1}$ bulk velocity and 10$^{-4}$~M$_\odot$ total mass. Mixing with
the underlying white dwarf material affects the chemistry of the ejecta,
which is dominated by non-equilibrium CNO-burning products. Classical novae
are instrinsically very luminous objects (with a luminosity approaching or
exceeding the Eddington luminosity for a 1~M$_\odot$ object, and $M_V$
staying brighter than $-$5.5~mag for the first two weeks past maximum), and
are easily observed throughout a vast fraction of our Galaxy, the Local
Group, and beyond it up to the Virgo Cluster. The well-established relation
between their absolute magnitude and the speed of decline makes novae useful
distance indicators (e.g. Cohen 1985).

Significant research carried out in the past at the time of photographic
spectroscopy to qualitatively characterize the spectroscopic evolution of
novae (e.g. Payne-Gaposchkin 1957, hereafter PG57, and references therein)
has led to the discovery of common paths followed by the majority of novae,
irrespective of their speed class and parentship with galactic stellar
populations (eg. McLaughlin 1960, hereafter McL60). Analogous and more
quantitative investigations in the era of CCD detectors are appreciably
rarer due to the large amount of telescope time allocation required,
expecially if observations aim to cover the entire nova outburst. Similarly,
historical photographic or amateur visual lightcurves exist for many novae
(e.g. Mobberley 1999, Kiyota 2004), but densely mapped, multi-band, and
accurate CCD lightcurves extending from maximum to the faintest evolutionary
stages are quite rare.

  \begin{table}[!h]
     \caption{Portion of Table~1, available only electronically, to illustrate its 
              content. The quoted error is the Poisson error and the last column
              provides identification of the telescope (see details in Sect.~2.1).}
     \centering
     \includegraphics[width=7.0cm]{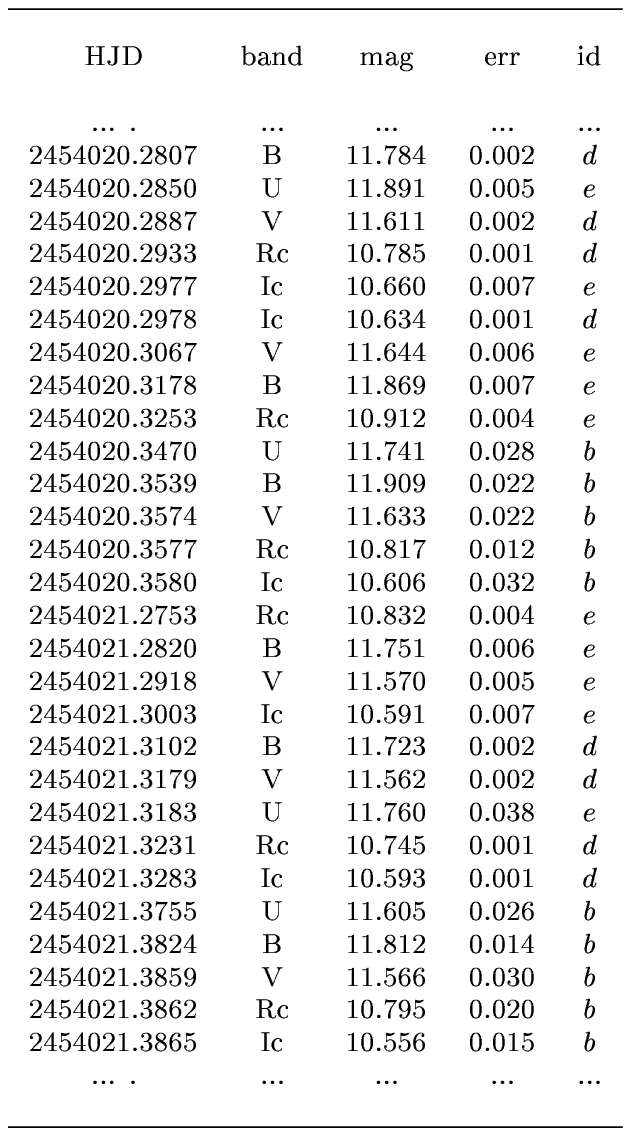}
     \label{tab1}
  \end{table}

This paper presents new and extensive multiwavelength photometric and
spectroscopic observations of Nova Cyg 2006 (= V2362 Cyg, hereafter NCyg06).
It was discovered at magnitude 10.5 by H. Nishimura on photographic plates
exposed on April 2.807 UT (as reported by Nakano 2006), while the nova was
still on the rise to its maximum. Within a few hours of the discovery
announcement, several spectroscopic confirmations of the nature of the nova
were made, as summarized by Yamaoka (2006a). This nova is characterized by one
of the most unusual photometric behaviors ever recorded for a classical
nova.  We investigate in detail the initial outburst, the rebrightening that
occurred several months later, and the advanced decline phase. We also
report photo-ionization modeling of the emission spectrum from the
expanding ejecta to estimate their total mass, their chemical partition, and
the temperature and luminosity of the ionizing central white dwarf, thus
adding to the limited number of novae ($\sim$30) for which such
analysis has been performed.
  
  \begin{figure}[!h]
     \centering
     \includegraphics[width=7.0cm]{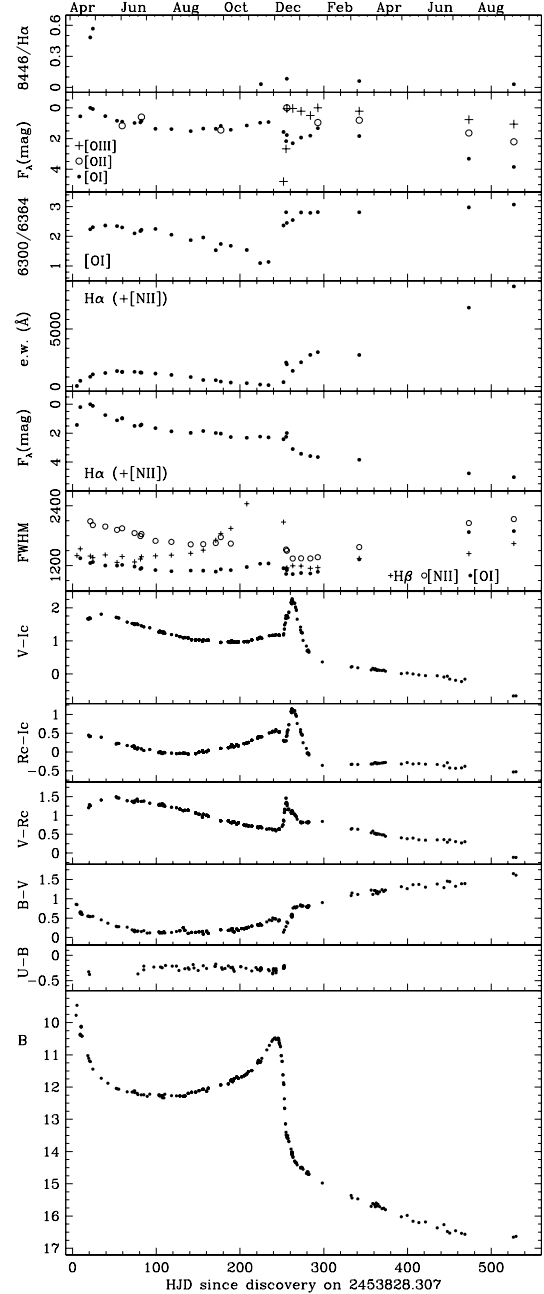}
     \caption{{\em Lower six panels}: photometric evolution of Nova Cyg 2006
              in the $U$$B$$V$$R_{\rm C}$$I_{\rm C}$ bands. {\em Upper six
              panels}: temporal evolution of [OI] 8446/H$\alpha$ and [OI]
              6300/[OI] 6364 flux ratios, of [OI] 6300, [OII] 7325, [OIII]
              5007, and H$\alpha$ integrated line flux (expressed in
              magnitudes relative their peak values 9.364$\times$10$^{-12}$,
              7.151$\times$10$^{-12}$, 1.249$\times$10$^{-11}$, and
              3.351$\times$10$^{-10}$ erg~cm$^{-2}$~sec$^{-1}$,
              respectively), of H$\alpha$ equivalent width, and of FWHM (in
              km/s) of H$\beta$, [OI]~6300, and [NII]~5750~\AA\ emission line
              profiles.  After day +250, a growing fraction of [NII] 6548,
              6584 emission contributed to the H$\alpha$ flux and equivalent
              width.}
     \label{fig1}
  \end{figure}

\section{Observations}

\subsection{Photometry}

We collected (cf. Table~1) CCD and photoelectric photometry of NCyg06
with a number of telescopes, densely covering all stages of its temporal
evolution. All observations were reduced against the $U$$B$$V$$R_{\rm
C}$$I_{\rm C}$ comparison sequence presented by Frigo et al. (2006). This
sequence spans a wide range in color and allows suitable transformation from
the local to the standard systems. The different instruments are associated
in the next paragraph to letters $a$ to $g$, which are used in Table~1 and
throughout the rest of the paper to identify them.
 
Observations were obtained with: ($a$) the Sonoita Research Observatory
(SRO) 0.35-m Celestron C14 robotic telescope using $B$$V$$R_{\rm C}$$I_{\rm
C}$ Optec filters and an SBIG STL-1001E CCD camera, 1024$\times$1024 array,
24 $\mu$m pixels $\equiv$ 1.25$^{\prime\prime}$/pix, with a field of view of
20$^\prime$$\times$20$^\prime$; ($b$) the 0.42-m f/5.4 Newtonian telescope
operated by Associazione Ravennate Astrofili Rheyta in Bastia (Ravenna,
Italy), equipped with an Apogee Alta 260e CCD camera, 512$\times$512 array,
20 $\mu$m pixels $\equiv$ 1.83$^{\prime\prime}$/pix, field of view of
16$^\prime$$\times$16$^\prime$ and Schuler $U$$B$$V$$R_{\rm C}$$I_{\rm C}$
filters; ($c$) the 0.50-m f/8 Ritchey-Chr\'etien telescope operated on top of
Mt. Zugna by Museo Civico di Rovereto (Trento, Italy) and equipped with
Optec $U$$B$$V$$R_{\rm C}$$I_{\rm C}$ filters. The CCD is an Apogee Alta U42
2048$\times$2048 array, 13.5 $\mu$m pixels $\equiv$
0.70$^{\prime\prime}$/pix, with a field of view of 24$^\prime$$\times$24$^\prime$; 
($d$) the 0.30-m Meade RCX-400 f/8
Schmidt-Cassegrain telescope owned by the Associazione Astrofili Valle di Cembra
(Trento, Italy). The CCD is an SBIG ST-9, 512$\times$512 array, 20 $\mu$m
pixels $\equiv$ 1.72$^{\prime\prime}$/pix, with a field of view of
13$^\prime$$\times$13$^\prime$. The $B$ filter is from Omega and the $V$$R_{\rm
C}$$I_{\rm C}$ filters from Custom Scientific; ($e$) the 0.13-m f/6.6 Vixen
ED130SS refractor privately owned by one of us (G.C.) and operated in
Trieste (Italy). It is equipped with Custom Scientific $U$$B$$V$$R_{\rm
C}$$I_{\rm C}$ filters and a Starlight SXV-H9 CCD camera, 1392$\times$1040
array, 6.45 $\mu$m pixel $\equiv$ 1.55$^{\prime\prime}$/pix for a field of
view of 36$^\prime$$\times$27$^\prime$; ($f$) the 0.28-m Celestron C11
telescope privately owned by one of us (S.D.) and operated in Cembra
(Trento, Italy). It is used with an Optec SSP-5 photoelectric photometer and
standard $B$,$V$ Johnson filters; and finally ($g$) the 1.0-m
Ritchey-Chr\'etien telescope of the U. S. Naval Observatory, Flagstaff
Station (NOFS). A Tektronix/SITe 2048x2048 thinned, backside--illuminated
CCD was used. The telescope scale is 0.6763 arcsec/pixel, with a field of
view of 23$^\prime$.1$\times$23$^\prime$.1 arcmin.

The 2353 photometric measures (89  in $U$, 614 in $B$, 621 in $V$, 498 in
$R_{\rm C}$, and 531 in $I_{\rm C}$) we collected on NCyg06 on 313 different
nights are listed in Table~1 (available electronic only). The overall nova
light-curve and color evolution are shown in Fig.~1 (see also Kimeswenger et
al. 2008), using data only from instruments $a$, $d$, and $g$
(Figs.16-20, available electronic only, present the equivalent plots
using all of the data in Table~1).

  \begin{table}
     \caption{Journal of spectroscopic observations.}
     \centering
     \includegraphics[width=8.8cm]{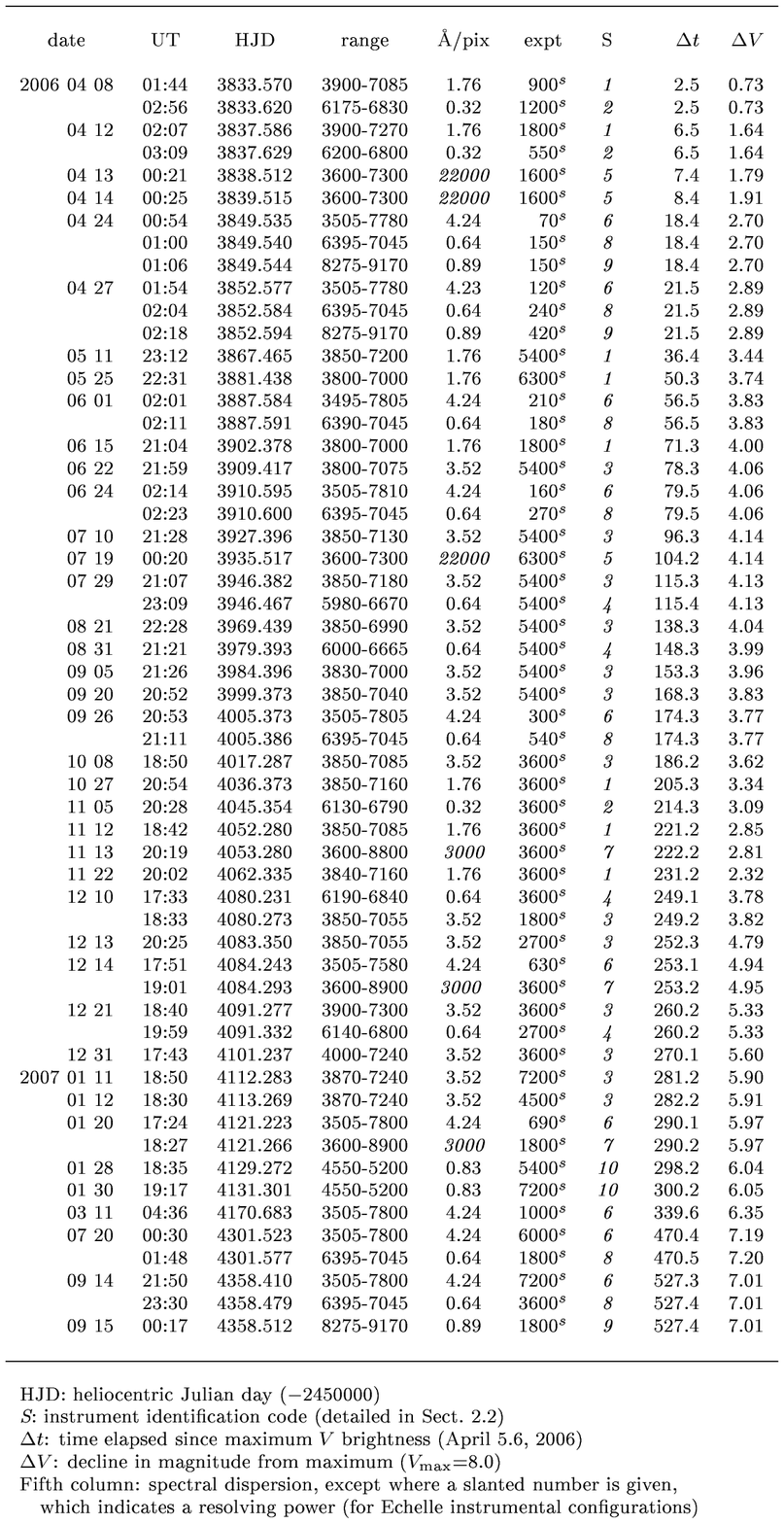}
     \label{tab2}
  \end{table}

\subsection{Spectroscopy}

A journal of the spectroscopic observations we obtained on NCyg06 is given
in Table~2, where a number from {\em 1} to {\em 10} in column seven
identifies the different telescope/spectrograph configurations, described
below. Reduction of all spectroscopic observations was carried out in a
homogeneous way with IRAF, and involved corrections for bias, dark, and flat
fields. All spectra (with the exception of the Echelle ones) were
calibrated into absolute fluxes via observations on the same nights and
instrumental configurations of one or more of the following
spectrophotometric standard stars: HR 718, HR 5501, 7001, 8634, and HD
161817. On several nights, the non-photometric conditions required a shift
be applied to the zeropoint of the fluxed spectra of the nova. This
shift was determined by comparing the fluxes integrated
over the photometric bands on the nova spectra with the corresponding fluxes
from CCD photometry of the nova for the same night (or extrapolated from
adjacent nights in one case). Higher resolution observations of the
nova (codes {\em 2}, {\em 4}, {\em 5}, {\em 7}, {\em 8}, {\em 9}, and {\em
10} in Table~2) were always accompanied by identical observations on the
same night of one or more of the following stars: HD 196740, HD 209833, HD
212571. These are fast-rotating B stars with no H$\alpha$ emission, and
their flat, featureless continua provided an excellent reference background
to record interfering absorptions by telluric H$_2$O and O$_2$.

Low and medium resolution, absolutely fluxed spectra of NCyg06 were obtained
with the 0.6-m telescope of Osservatorio Astronomico G. Schiaparelli
(Varese, Italy), equipped with a grating spectrograph and an SBIG ST-10XME CCD,
2184$\times$1472 array, 6.8 $\mu$m pixel. The slit was always aligned east-west,
and its projected width was kept constant to 2.0$^{\prime\prime}$
through the whole observing campaign. The spectrograph spatial scale is
1.0~$^{\prime\prime}$/pix. Five different set-ups were used: 600 ln/mm
grating, range 3900-7100~\AA, 1$\times$1 binning, and 1.76~\AA/pix (code {\em
1} in Table~2); 600 ln/mm grating, range 3900-7100~\AA, 2$\times$2 binning,
and 3.52~\AA/pix (code {\em 3}); 1800 ln/mm grating, range 6200-6900~\AA,
1$\times$1 binning and 0.32~\AA/pix (code {\em 2}); 1800 ln/mm grating,
range 6200-6900~\AA, 2$\times$2 binning, and 0.64~\AA/pix (code {\em 4}); and
1800 ln/mm grating, range 4550-5200~\AA, 2$\times$2 binning, and 0.83~\AA/pix
(code {\em 10}).
 
Low and medium resolution, absolutely fluxed spectra of NCyg06 were also obtained
with the AFOSC imager+spectrograph mounted on the 1.82m telescope
operated in Asiago by the INAF Astronomical Observatory of Padova. It is
equipped with a Tektronix TK1024 thinned CCD, 1024$\times$1024 array, 24
$\mu$m pixel, with a scale perpendicular to dispersion of
0.67~$^{\prime\prime}$/pix. All observations were obtained with a
1.26$^{\prime\prime}$ slit always aligned east-west.  Four different
instrument set-ups were used: 300 ln/mm grism, range 3500-7780~\AA,
1$\times$1 binning, and 4.24~\AA/pix (code {\em 6} in Table~2); 1720 ln/mm
volume phase holographic grism, range 6400-7050~\AA, 1$\times$1 binning, and
0.64~\AA/pix (code {\em 8}); 1280 ln/mm volume phase holographic grism,
range 8270-9270~\AA, 1$\times$1 binning, and 0.89~\AA/pix (code {\em 9}); and
and the combination of a 79 ln/mm Echelle grism + 150 ln/mm cross-disperser
grism covering the whole wavelength range 3600-8800~\AA\ in 13 orders with no gaps, 
providing a 3\,000 resolving power with 1$\times$1 binning
and 2.4 $^{\prime\prime}$ slit width (code {\em 7}).

High resolution spectra were obtained with the Echelle spectrograph mounted
on the 1.82m Asiago telescope (code {\em 5} in Table~2). It is equipped with
an EEV~CCD47-10 CCD, 1024$\times$1024 array, 13 $\mu$m pixel, covering the
interval $\lambda\lambda$~3600$-$7300~\AA\ in 32 orders (wavelength gaps
between adjacent orders are present redward of 4900~\AA). The slit was
always aligned east-west, and its 2$^{\prime\prime}$ projected width
provided a 20\,000 resolving power.

  \begin{figure}
     \centering
     \includegraphics[width=8.8cm]{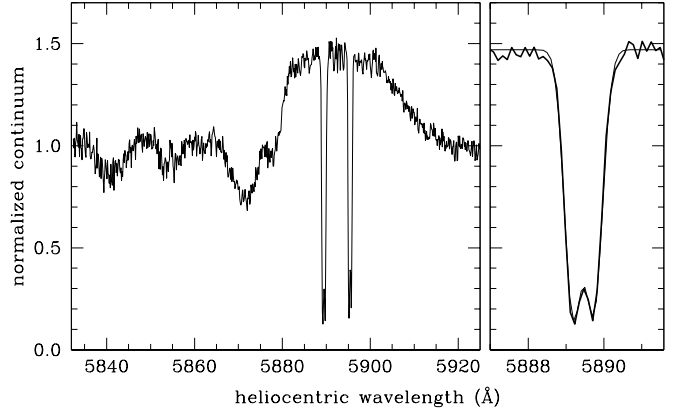}
     \caption{{\em Left panel}: a small portion of the combined Echelle spectra
              for April 13 and 14, 2006 of Nova Cyg 2006,
              centered on the interstellar NaI D$_{1,2}$ absorption lines.
              {\em Right panel}:
              Blowup of the interstellar NaI D$_2$ line (thicker line)
              showing the fit with two components at $-$39.1 ($\pm$0.4) and
              $-$11.1 ($\pm$0.3)~km~sec$^{-1}$ (thinner line).}
     \label{fig2}
  \end{figure}

\section{Astrometry and progenitor}

Astrometry was performed on SRO images using SLALIB (Wallace 1994) linear plate
transformation routines in conjunction with the UCAC2
reference catalog. Errors in coordinates were typically under 0.1 arcsec in both
coordinates, referring to the mean coordinate zero point of the reference
stars in each field. The coordinates we derived for NCyg06 are\\
\begin{eqnarray}  \alpha_{\rm J2000}  &=&~~21^h 11^m 32^s.33 
                  ~(\pm 0^{\prime\prime}.12) \nonumber \\
                  \delta_{\rm J2000}  &=&+44^\circ 48^\prime 03^{\prime\prime}.74 
                  ~(\pm 0^{\prime\prime}.03), \nonumber
\end{eqnarray}

\noindent
close to the coordinates measured by Yamaoka (2006b) at
32$^s$.35 ($\pm$0$^{\prime\prime}$.15) and 03$^{\prime\prime}$.66
($\pm$0$^{\prime\prime}$.14). The corresponding galactic coordinates are
$l$=087$^\circ$.372 and $b$=$-$02$^\circ$.357.

At this position, Steeghs et al. (2006) identified a faint point source on
IPHAS images for Aug 3, 2004, measured at $r^\prime$=20.30 ($\pm$0.05) and
$i^\prime$=19.76 ($\pm$0.07) and with appreciable H$\alpha$ emission, which
they argue is the progenitor of the nova. A search of archival 
Asiago Schmidt plates for 1962-1984 was carried out by Jurdana-Sepi\v{c} and 
Munari (2006) who found no brightening above the $B$$\sim$18.5 plate limit. 
Both POSS-I and -II images also do not show a counterpart at the position of 
the nova.

\subsection{New Asiago archive plates}

In addition to the photographic plates already measured by Jurdana-Sepi\v{c}
and Munari (2006), we identified in the plate archives of the Asiago
Schmidt telescopes another block of 57 direct-image plates covering the
position of the nova progenitor over the period 1989-1997. All them have
been inspected to search for a possible brightening during quiescence that
could have caused the progenitor to rise above the $B$$\sim$18.5 plate
limit. None has been found. Table~3 (available electronic only) reports
details on these plates.

To estimate the $B$-band magnitude expected in quiescence for the nova
progenitor, we adopted the mean fluxed spectrum from Zwitter and Munari (1995) 
for a cataclysmic variable in quiescence, increased its reddening to
match the $E_{B-V}$=0.56 affecting NCyg06 (see Sect. 4), and scaled its flux
to match the Steeghs et al. (2006) $r^\prime$ and $i^\prime$ IPHAS
magnitudes. This caclulation indicates that the progenitor should be
$B$$\sim$21.0 mag in quiescence. The negative detections on Asiago archive
plates therefore exclude CV-type outbursts with an amplitude greater than
$\Delta$$B$$\sim$2.5~mag. Frequent and large amplitude outburst activity
as displayed by SS Cyg ($<$$\Delta V$$>$$\approx$4~mag, outbursts every
$\approx$55 day, of $\approx$15 day duration; cf. Cannizzo and Mattei 1992),
a prototype cataclysmic variable, should have been easily detected on
the Asiago archive plates.

  \begin{figure}
     \centering
     \includegraphics[width=8.8cm]{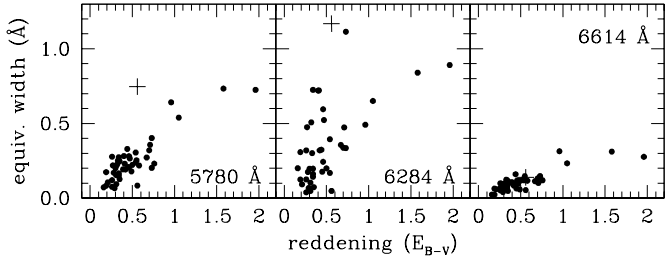}
     \caption{Comparison between the equivalent width of three diffuse
     interstellar bands and interstellar reddening for field stars measured
     by Megier et al. (2005). The crosses represent the position of Nova Cyg
     2006.}
     \label{fig3}
  \end{figure}

\subsection{The faint optical companion}

Seven arcsec west of the nova lies a faint field star, identified with 2MASS
21113188+4448032 ($J$=13.716 $\pm$ 0.029, $H$=13.441 $\pm$ 0.046,
$K_s$=13.480 $\pm$ 0.037). We measured on NOFS images for JD=2454247
its optical magnitude and colors as $V$=15.22, $U-B$=+0.33, $B-V$=+0.70,
$V-R_{\rm C}$=+0.42, $R_{\rm C} - I_{\rm C}$=+0.46, $V - I_{\rm C}$=+0.88,
with errors not exceeding 0.02 mag. Its astrometric position (J2000) from
these images, with respect to the UCAC2 reference catalog, is $\alpha$=21 11
31.88 ($\pm$ 0.1 arcsec), $\delta$=+44 48 03.2 ($\pm$ 0.1 arcsec). Steeghs
et al. (2006) measured $r^\prime$=14.8 for this star, in line with our
$R_{\rm C}$=14.80.

The faint optical companion could not be resolved until the advanced decline
set in $\sim$260 days past maximum, when its contribution to the measurement
of NCyg06 was no longer negligible. At later dates, the companion was
spatially resolved only by instruments $a$ and $g$ thanks to better seeing
and sampling. Photometry in Table~1 obtained with instruments $b$, $c$, $d$,
$e$, and $f$ always includes the contribution by the faint optical companion,
which can be easily accounted for by using the companion's $U$$B$$V$$R_{\rm
C}$$I_{\rm C}$ magnitudes. The photometry in Table~1 obtained with
instruments $a$ and $g$ only includes the contribution by the faint optical
companion for $t$$<$260. At later dates, its was possible to measure
the companion and the nova via PSF-fitting separately, and the values
reported in Table~1 for $t$$\geq$260 and instruments $a$ and $g$ pertain to
the nova alone. All values quoted in the text and the light- and
color-curves presented in this paper (Figs.~1, 4, 10, 15-20) have the
contribution by the companion removed.

\section{Reddening}

Sharp NaI D$_{1,2}$ interstellar absorption lines are prominent on the 
high-resolution Echelle spectra of NCyg06 obtained on April 13 and 14, 2006,
close to the optical maximum. Figure~2 shows the overall appearance of the
interstellar lines superimposed onto the emission component of the wide NaI
P-Cyg profile of the nova itself, as well as an enlargement of their profile
that reveals how the interstellar lines are actually a blend of two distinct
components of similar intensity. Their heliocentric radial velocities are
$-$39.1~($\pm$0.4) and $-$11.1~($\pm$0.3)~km~sec$^{-1}$, and their
equivalent widths (for the D2 component at 5890 \AA) are 0.484 and
0.487~\AA, respectively. Following the relation between the interstellar
reddening and equivalent width of NaI D2 line calibrated by Munari and
Zwitter (1997), both components correspond to a reddening of 0.28 mag, for a
total reddening of $E_{B-V}^{TOT}$= 0.28 + 0.28 = 0.56, which will be
adopted throughout this paper.

van den Bergh and Younger (1987) derived a mean intrinsic color
$(B-V)_\circ$=$-$0.02 $\pm$0.04 for novae at $t_2$. Figure~1 and the
photometry in Table~1 show that NCyg06 at $t_2$ had $B-V$=0.54. This
corresponds to $E_{B-V}$=0.56, a perfect match to the result from
interstellar lines. Russell et al. (2006) preliminarily estimated
$E_{B-V}$=0.59 from infrared spectroscopy.

There are several diffuse interstellar bands (DIBs) visible on the NCyg06
spectra. On the April 8 spectra (when the continuum was strongest and
smoothest), DIBs at 6203, 6270, and 6376-79 \AA\ are present but too weak for
a meaningful measure against the overlapping P-Cyg components of FeII
multiplet 74 lines. DIBs at 4430, 5780, 6284, and 6614~\AA\ show equivalent
widths of 2.57, 0.75, 1.17, and 0.14~\AA, respectively. A comparison of the
equivalent width of the last three DIBs with the reddening/equivalent widths
measured for field stars by Megier et al. (2005) is presented in Fig.~3. The
correlation between reddening and DIB equivalent widths is known to be a
loose one (cf Herbig 1995). Nevertheless, it is worth noticing that, while
the intensity of DIB 6614~\AA\ in NCyg06 seems to nicely follow the mean
relation for field stars, the DIBs at 5780 and 6284 are appreciably stronger
than in field stars. DIB~4430~\AA\ similarly lies above the mean relation
for field stars (though still within the scatter of individual data)
presented by Isobe et al. (1986). Overall, the DIBs toward NCyg06 are
somewhat stronger than expected for the $E_{B-V}$=0.56 reddening affecting
the nova. The nova itself should not be contributing to the observed DIBs
because their very low observed heliocentric radial velocities are what is
expected for an origin in the interstellar medium and bear no relationship
to the high radial velocity of the ejecta.

\section{Early evolution}

The early evolution of NCyg06 was that of a fairly normal FeII-type nova,
from both the spectroscopic and photometric points of view, with no hint of
the later peculiarities that led to one of the most peculiar lightcurves
ever recorded for a nova.

\subsection{Photometry}

The first two months of the nova light-curve are presented in
Fig.~4. The optical photometric evolution of NCyg06 was characterized by a
fast rise to a maximum of $V$=8.0 ($\pm$0.1) on Apr 5.6 ($\pm$0.5) 2006
UT (=JD 2453831.1). The time required to raise the final two magnitudes was
$t_{\rm r,2}$$\sim$2.3~days, and the nova was discovered 2.8 days before maximum,
when it was 2.5 mag fainter. A week before maximum, the nova was below
the patrol plate limit at 12 mag (cf IAUC 8698). The decline in $V$-band during
the first two months (4$<$$\Delta t$$<$56, where $\Delta t$=$t$-$t$$_{\rm max}$
is the time in days from maximum) strictly followed 
the exponential slope given by
\begin{equation}
V(t) = 7.6 + 2.4\times \log (\Delta t)
\end{equation}
shown in Fig.~4 as a solid curve. A large ($\Delta B$$\approx$0.3 mag)
variability of NCyg06, but with constant color ($\Delta B-V$$\approx$0.0),
is superimposed on the general decline during the first ten days after
maximum. This scatter cannot be ascribed to observational problems. In fact,
the corresponding observations have the highest accuracy, and the results
are confirmed from independent instruments, as well as repeated observations 
with the same instrument.

The nova decline times were $t_{2}^{V}$=10.4 and $t_{3}^{V}$=24 days, on the
borderline between {\em very fast} and {\em fast} novae. Both $t_{2}$ and $t_{3}$
are in normal proportion among them. In fact, the following relation 
fits nicely the sample of $\sim$20 best-studied novae away from the
galactic Bulge (i.e. 45$^\circ$$\leq$galactic longitude$\leq$305$^\circ$),
\begin{equation}
t_{3} \sim 6\times t_{2}^{0.7} - 7,
\end{equation}
and NCyg06 falls right on it. The rise $t_{\rm r,2}$ and decline $t_{2}$,
$t_{3}$ times of NCyg06 are close to the proportionality relation found by
Schmidt (1957): $\log$~$t_{\rm r,2}$ = $-$0.3 + 0.7~$\log$~$t_{2}$ = $-$0.5
+ 0.7~$\log$~$t_{3}$. In addition, the $\sim$13.3~mag outburst amplitude and
$t_{2}^{V}$=10.4 fall precisely on the Warner (1995) relation between
amplitude and $t_{2}$.

  \begin{figure}
     \centering
     \includegraphics[width=8.8cm]{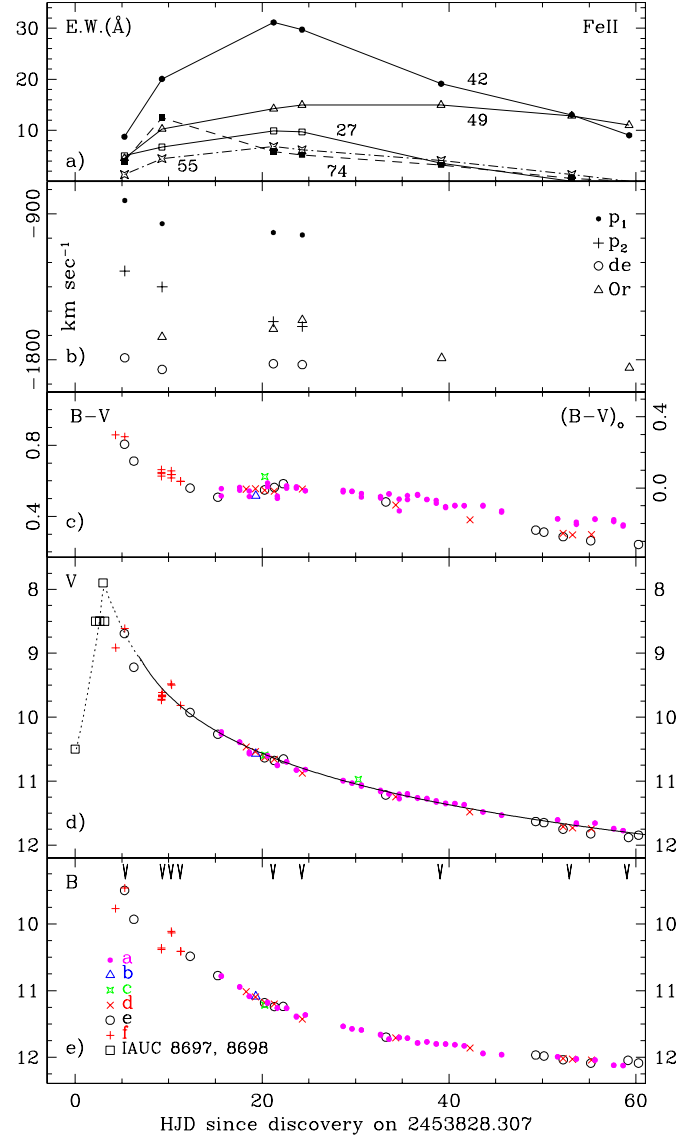}
     \caption{The early photometric decline of Nova Cyg 2006. For instrument
              identification see Sect 2.1. The solid curve over-plotted to
              the $V$ data in panel $d$ is Eq. (1), while the dotted curve
              is drawn by hand only for eye guidance. The $a$ panel shows
              the evolution in equivalent width of an un-blended emission
              line for each of the FeII multiplets 27, 42, 49, 55 and 74
              (4233, 4924, 5317, 5535 and 6239 ~\AA, respectively). Panel
              $b$ presents the change in radial velocity of absorption
              systems (the {\em "de"} diffuse-enhanced and the $"p_1"$ and
              $"p_2"$ principal systems seen in H$\alpha$, and the {\em
              "Or"} Orion system seen in HeI 6678~\AA).}
     \label{fig4}
  \end{figure}

van den Bergh and Younger (1987) found that the average intrinsic color of
novae at maximum is $(B-V)_\circ$=+0.23 $\pm$0.06 (dispersion 0.16 mag). Our
first color measurement was obtained close to maximum at $\Delta t$=+1$^d$.6
(cf. Table~1). It gave $B-V$=+0.85, which corrected for the $E_{B-V}$=0.56
reddening, corresponds to $(B-V)_\circ$=+0.29, thus reasonably close to the
mean of intrinsic colors of novae at maximum.

  \begin{figure*}
     \centering
     \includegraphics[width=18.0cm]{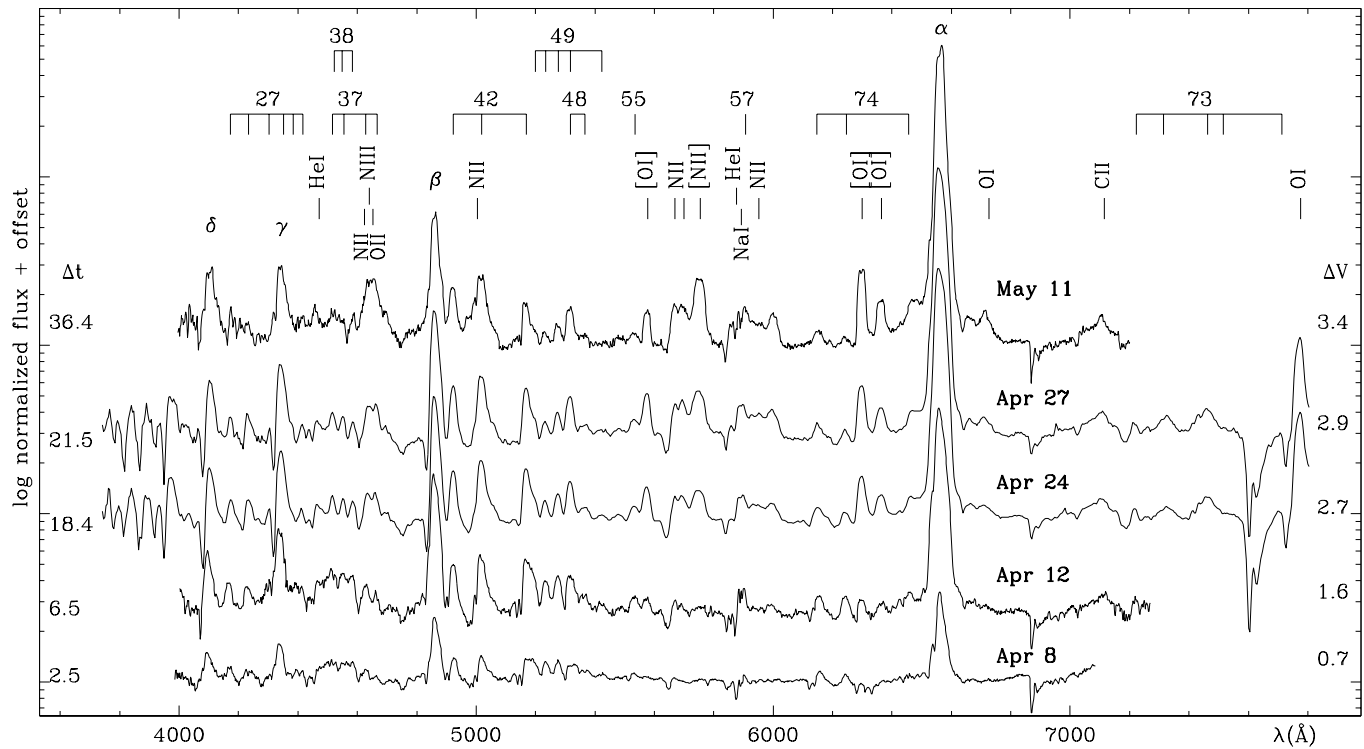}
     \caption{Early spectroscopic evolution of Nova Cyg 2006. The spectra are
              continuum-normalized and shifted to avoid overlap. The ordinates are
              in logarithm of the flux to emphasize the visibility of weaker features. 
              $\Delta t$ and $\Delta V$ are days and magnitudes from maximum.
              Emission lines are identified (FeII multiplets by
              their numbers and comb-like markings).}
     \label{fig5}
  \end{figure*}

  \begin{figure}
     \centering
     \includegraphics[width=6.8cm]{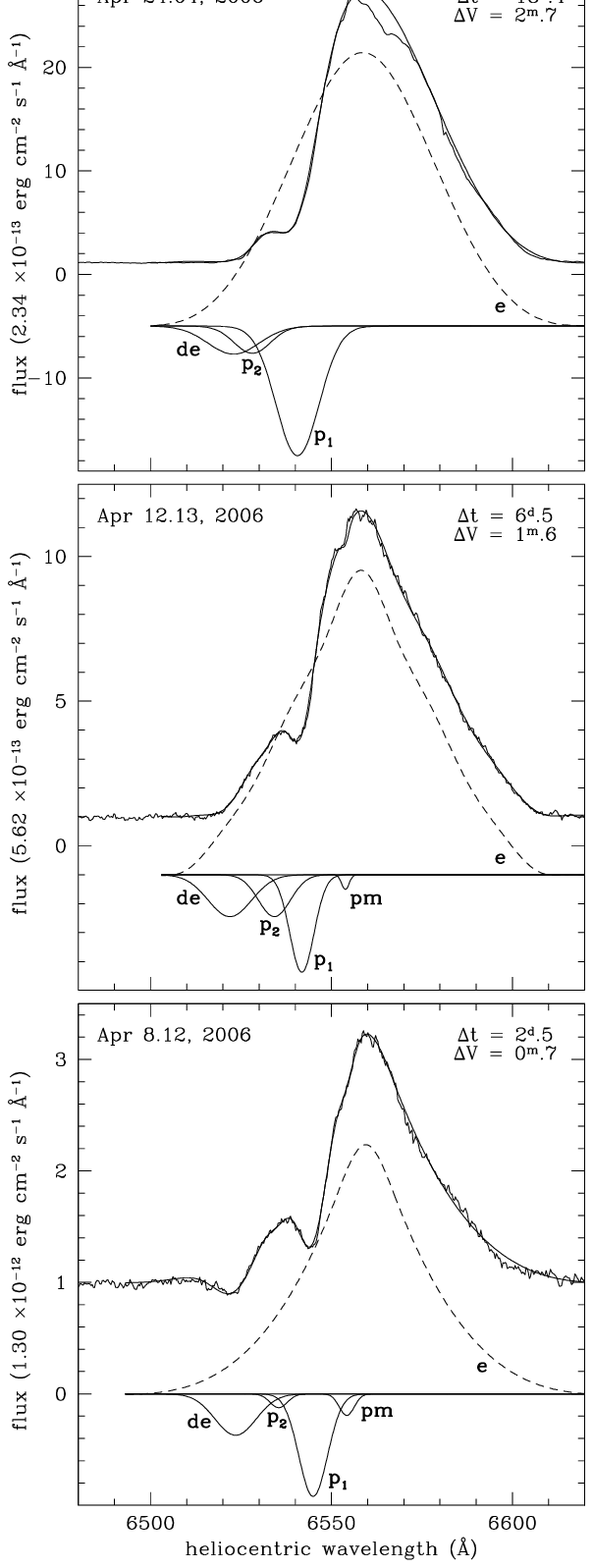}
     \caption{Evolution of the H$\alpha$ profile at three epochs during the 
              first three weeks past maximum (see Sect 5.3.1 for details).}
     \label{fig6}
  \end{figure}

\subsection{Distance and bolometric magnitude}

Both the rate of decline and the observed magnitude 15 days past maximum are
popular methods of estimating the distance to the nova.

Published relations between absolute magnitude and the rate of decline
generally take the form $M_{\rm max}\,=\,\alpha_n\,\log\, t_n \, + \,
\beta_n$. Given the lack of observations in $B$ band right at maximum light,
we are limited to equations valid for $V$ band alone.  Using the Cohen (1988)
$V$-$t_2$ relation, the distance to NCyg06 is 8.0~kpc, and it is 7.3~kpc according
to the Schmidt (1957) $V$-$t_3$ relation. Using the della Valle and Livio
(1995) expression calibrated on novae appearing in the LMC and M31, the
distance to NCyg06 would increase to 9.7~kpc. Since it is not applicable to
galactic novae, it will be not considered further.

Buscombe and de Vaucouleurs (1955) suggested that all novae have the same
absolute magnitude 15 days after maximum light. There are several
calibrations of it:
$M_{15}^{V} $=$-$5.2$\pm$0.1 from Buscombe and de Vaucouleurs (1955),
$M_{15}^{V} $=$-$5.60$\pm$0.43 from Cohen (1985),
$M_{15}^{V} $=$-$5.23$\pm$0.16 from van den Bergh and Younger (1987),
$M_{15}^{V} $=$-$5.38 from van den Bergh (1988),
$M_{15}^{V} $=$-$5.69$\pm$0.14 from Capaccioli et al. (1989),
$M_{15}^{pg}$=$-$5.85 according to Schmidt (1957),
$M_{15}^{pg}$=$-$5.50$\pm$0.18 from de Vaucouleurs (1978), and 
$M_{15}^{B} $=$-$5.74$\pm$0.60 from Pfau (1976). The brightness of NCyg06
15 days after maximum light was $V_{\rm 15}$=10.45, $B_{\rm 15}$=10.97.
Comparing with the above absolute magnitudes and correcting for $E_{B-V}$=0.56, the
distance turns out to be 6.1, 7.3, 6.2, 6.6, 7.6, 8.0, 6.8, and 7.6,
respectively (approximating $m_{pg}$$\approx$$B$).

Taking the unweighted mean of the ten determinations above, the distance to
NCyg06 is $d$=7.2~kpc, with an error of the mean of 0.2~kpc and a dispersion
$\sigma_d$ = 0.65~kpc.  The bolometric correction for an expanded
photosphere with $B-V$=+0.29 is {\em B.C.}=$-$0.02 (from Drilling and
Landolt 2000). Coupled with a 7.2~kpc distance, $E_{B-V}$=0.56 and $V_{\rm
max}$=8.0 mag, it gives for NCyg06 a bolometric magnitude at maximum of
$M_{\rm bol}$=$-$8.0~mag.

At NCyg06 galactic latitude, a 7.2~kpc distance translates to a height above
the galactic plane of $z$=0.3~kpc, still within the vertical scale height of
the galactic Thin Disk. At galactic longitude $l$=87$^\circ$.37, the line of
sight to NCyg06 crosses the Perseus spiral arm 5~kpc away from the Sun, and
stops midway between the Perseus and Cygnus spiral arms (cf. Vall\'{e}e
2005). According to the Brand and Blitz (1993) maps, the radial velocity of
interstellar matter belonging to the Perseus spiral arm where it is crossed
by the line of sight to NCyg06 is $\approx$$-$40~km~sec$^{-1}$, so the
$-$39~km~sec$^{-1}$ component seen in the high-resolution profiles of
interstellar NaI~D$_{1,2}$ (Fig.~2 and Sect.~4) is quite possibly associated
with the Perseus spiral arm. The other component at $-$11~km~sec$^{-1}$
should be connected to material in front of the Perseus spiral arm, as
suggested by the Neckel and Klare (1980) extinction maps and by the lower
observed radial velocity.

  \setcounter{table}{3}
  \begin{table}
     \caption{Heliocentric velocity, velocity span at half maximum,
              equivalent width, and integrated absolute flux (in units of
              10$^{-13}$ erg cm$^{-2}$ sec$^{-1}$) of the emission
              component and pre-maximum, principal and diffuse enhanced
              absorption systems for the H$\alpha$ profiles shown in
              Fig.~6 (and in addition for the H$\alpha$ profile of 
              Apr 27.09, 2006 not shown in Fig.~6 to save space).}
     \centering
     \includegraphics[width=6.8cm]{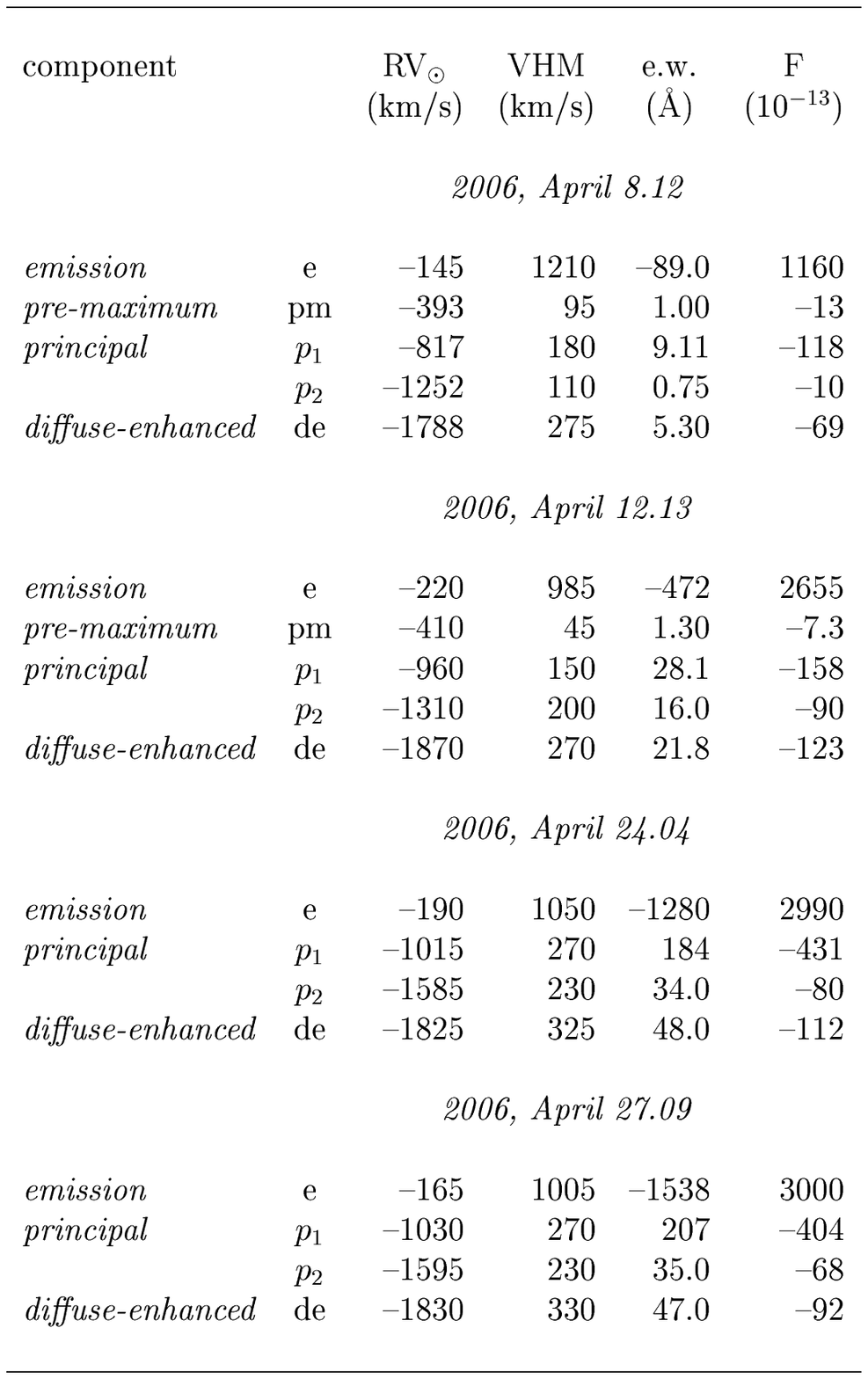}
     \label{tab4}
  \end{table}

\subsection{Spectroscopy}

The low-resolution, broad-wavelength range spectroscopic evolution of NCyg06
during the first two months of the outburst is presented in Fig.~5.

\subsubsection{Absorption systems}

The evolution of the H$\alpha$ profile during the first three weeks of the
outburst is shown in Fig.~6. It is based on high-resolution observations
obtained at epochs $\Delta t$=+2.5, +6.5, and +18.4 days. A fourth profile for
$\Delta t$=+21.5, similar to that of day +18.4, is not shown to save space,
but it is nevertheless included in Table~2, which reports the results of the
H$\alpha$ profile deconvolution at all four epochs. We do not consider
later observations because of the reinforcement of [NII] emission in 
May and the blending of its 6548, 6584 \AA\ lines with the H$\alpha$.

The H$\alpha$ profile fitting was carried out assuming that the
underlying emission component has a symmetric profile and that absorptions
have Gaussian profiles. The profile of the emission component is therefore
derived from the portion of the observed profile that extends from the peak
to the asymptotic merging of the red H$\alpha$ wing into the continuum. In
Fig.~6 the profile of the emission component is plotted with a dashed
line, those of the absorption components by solid lines.

The first spectrum ($\Delta t$=+2.5), obtained when the nova had already
declined by $\Delta V$=0.7~mag, shows four absorption components, listed
in Table~4. They correspond to the simultaneous presence of the
classical {\em pre-maximum}, {\em principal}, and {\em diffuse-enhanced}
absorption systems (indicated by $pm$, $p$, and $de$ in Fig.~6, respectively), 
which are common to most novae and are described in detail by McLaughlin (1960,
hereafter McL60). A noteworthy characteristic of NCyg06 is the splitting of
the {\em principal} absorption spectrum into two components ($p_1$, $p_2$),
as seen previously in other novae (e.g. N Aql 1918 and N Gem 1912). At
$\Delta V$=0.7~mag, the {\em pre-maximum} absorption system was already
declining in equivalent width, while the {\em principal} system was rapidly
gaining strength and the {\em diffuse-enhanced} system had just emerged.
Their heliocentric radial velocity, velocity width at half intensity and
integrated absolute flux are given in Table~4.

Data summarized by McL60 show a correlation between {\em mean} radial
velocity of the various absorption systems and the speed class of the nova
({\em mean} velocity because the velocity of any given system usually
changes with time). This correlation states $v_{\rm
pre-max}$$\approx$$-$4750/$t_2$ for the {\em pre-maximum} absorption system.
The $t_2$=10.4 days for NCyg06 would correspond to $-$450~km/s, close
to the observed $-$400~km/s (Table~4).  The velocity relation for the {\em
principal} system is $\log v_{\rm prin}$ = 3.70 $-$ 0.5$\log t_3$ = 3.57 $-$
0.5$\log t_2$, and predicts $\approx$$-$1080~km/s for NCyg06. The epoch
means of the $p_1$,$p_2$ velocities in Table~4 (weighted according to their
equivalent widths) are $-$850, $-$1087, $-$1103, and $-$1112 km~sec$^{-1}$.
Their global average is $-$1040~km~sec$^{-1}$, again close to
predictions. The McL60 relation for the {\em diffuse enhanced} system is
$\log v_{\rm dif-enh}$ = 3.81 $-$ 0.4$\log t_3$ = 3.71 $-$ 0.4$\log t_2$, and
it predicts $\approx$$-$1880~km/s velocity for NCyg06, in good agreement with the
averaged observed value of $-$1830~km/s.

All three absorption systems in NCyg06 appear to move $\sim$50~km~sec$^{-1}$
more slowly then predicted on the basis of the relations by McL60. The
difference is actually more than twice as large. In fact, the component of the
radial velocity due to the galactic rotation can be written as
\begin{eqnarray}
RV_\odot & = & w_\circ\ {\rm sin}b \ + \ u_\circ\ {\rm cos}l\ {\rm cos}b
               \ - \ v_\circ\ {\rm sin}l\ {\rm cos}b \nonumber \\
         &   & - \ 2[A(R - R_\circ) + \alpha (R-R_\circ)^2]\ {\rm sin}l\ {\rm cos}b
\end{eqnarray}
where $R^2 = R_{\circ}^{2} + d^2 - 2 R_\circ d {\rm cos} l$,
$A = - (R_\circ /2) (d\omega / dR)_{R_\circ}$ and $\alpha = - (R_\circ /
4) (d^2\omega / dR^2)_{R_\circ}$. Here, $R$ and $R_\circ$ are the nova and Sun 
galacto-centric distances, $d$ is the nova-Sun distance, ({\em l,b}) the nova 
galactic coordinates and ($u_\circ, v_\circ, w_\circ$) is the solar motion vector. 
Adopting ($u_\circ, v_\circ, w_\circ$)\ =\ (10.0,5.3,7.2) km sec$^{-1}$
from Dehnen and Binney (1998), and $A=17.0 \pm 1.5$ km sec$^{-1}$ kpc$^{-1}$ and
$\alpha = -2.0 \pm 0.6$ km sec$^{-1}$ kpc$^{-2}$ from Hron (1987), the
radial velocity of NCyg06 expected from galactic rotation would be
$-$69~km~sec$^{-1}$. Therefore, it seems that the mean {\em barycentric}
outflow velocity of the pre-maximum, principal and diffuse-enhanced spectra of
NCyg06 are about 120~km~sec$^{-1}$ slower than predicted by the McL60 relations.

The FeII line profiles in the low-resolution spectra for April 8 and 12 in
Fig.~5, and the Echelle spectra for April 13 and 14, also show the
presence of the principal and diffuse-enhanced absorption systems. The
pre-maximum system had become too weak to be detected on the FeII lines by
the time of the Echelle spectra. The radial velocities of the absorption
systems on the FeII lines were close to what was measured for H$\alpha$. On the
low-resolution April 8 spectrum, they were $\approx$$-$830 for $p_1$
principal and $\approx$$-$1700~km/s for diffuse-enhanced. On the Echelle
spectrum for April 12 they were $\approx$$-$985 for $p_1$ principal,
$\approx$$-$1295 for $p_2$ principal, and $\approx$$-$1780~km/s for
diffuse-enhanced. Also the ratio of the equivalent widths of $p_1$ and
diffuse-enhanced was similar in H$\alpha$ and FeII lines: 1.29 and 1.30,
respectively, on April 12. The only major difference with H$\alpha$ was in
the much weaker intensity in the FeII lines of the $p_2$ system compared
with the $p_1$ system.

The time behavior of the radial velocity of $p_1$, $p_2$ principal and
diffuse-enhanced absorptions on the H$\alpha$ profile is illustrated in
Fig.~4. The velocity of the $p_1$ system has been increasing with time as
$v_1$$\propto$$-215 \log (\Delta t)$, and for $p_2$ as $v_2$$\propto$$-400
\log (\Delta t)$~km/s. The velocity increase was minimal for the
pre-maximum absorptions, while there has been no increase at all in the
diffuse-enhanced systems. This agrees with what is seen in most
novae (cf. McL60).

Feeble traces of the emergence of the {\em Orion} absorption system appear
on the April 12 spectrum ($\Delta t$=+6.5, $\Delta V$=1.6 mag) of Fig.~5.
The Orion absorptions grew much stronger on the April 24 and 27 spectra
($\Delta t$=+20, $\Delta V$=2.8 mag). It was still weakly present on the
May~11 spectrum ($\Delta t$=+36, $\Delta V$=3.4 mag). The Orion system
affected the HeI, NII and OII lines but not the hydrogen ones, as is common
in many novae. The radial velocity and equivalent width of the Orion
absorption component of HeI 6678 \AA\ are listed in Table~5, while Fig.~4
shows its velocity progression in comparison with the other absorption
systems. In many novae, the velocity of the Orion absorption is comparable
to or higher than for the diffuse-enhanced system, while the
Orion system moves about 250~km/s slower in NCyg06.

  \begin{table}
     \caption{Radial velocity and equivalent width of Orion the absorption system
              as observed in the HeI 6678 \AA\ line.}
     \centering
     \includegraphics[width=4.8cm]{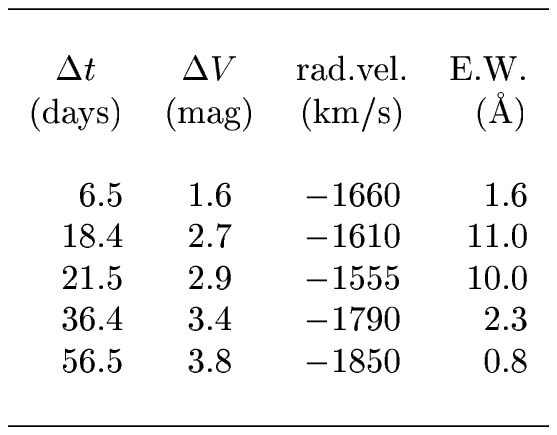}
     \label{tab5}
  \end{table}

In typical novae, as the numerous NII and OII lines of the Orion system fade
away, a pair of NIII absorption lines emerges at 4097 and 4103 \AA. This
also occurred in NCyg06, for which the NIII absorption is quite visible on
the blue wing of the H$\delta$ emission line of the June 1 spectrum ($\Delta
t$=+56, $\Delta V$=3.8 mag) illustrated in Fig.~9.  Its velocity is
$\approx$$-$2200~km~sec$^{-1}$. The June~1 spectrum corresponds to the date
when the nova abandoned the exponential decline of Eq.(1), began to slow its
decline rate toward the plateau phase of mid 2006 and then rose to the
second maximum of late 2006. This anomalous photometric behavior was
mirrored by an initially slowed, and then aborted, spectroscopic evolution.
In fact, by the time the NIII 4100~\AA\ absorption appears, a typical nova
is entering the transition phase, when its spectrum turns from the stellar
to the nebular type. This did not happen in NCyg06. It did not go through
the transition phase and the development of the NIII absorption soon
stopped. The equivalent width of the latter, in fact, reached a modest
1.1~\AA\ on the June~1 spectrum and did not rise to the expected
$\sim$10$\times$ larger equivalent width it should have attained at later
dates. Instead, it rapidly faded away, leaving no obvious trace on the
June~22 spectrum in Fig.~9.
 
\subsubsection{Emission lines}

Figure~5 shows the evolution of the emission lines during the first two
months of the outburst. On $\Delta t$=+2.5, the emission lines were
essentially restricted to the hydrogen Balmer series and emerging FeII lines
(principally from the multiplets 42, 49, and 74), on top of a poorly
structured, relatively cool continuum. The intrinsic color at that time,
(B-V)$_\circ$=+0.29, was still only marginally perturbed by emission lines
and corresponded to that of an A9/F0 III star. The emission line spectrum was
typical of the {\em FeII}-class defined by Williams (1992). The expansion
velocity (corrected for instrumental resolution) derived from the width at
half maximum was similar for all hydrogen Balmer emission lines and amounted
to 780~km~sec$^{-1}$; the velocity estimated from the width of FeII lines is
25\% slower. Four days later, on $\Delta t$=+6.5 the emission line spectrum
was getting much stronger (cf. the increase in equivalent width of H$\alpha$
and FeII lines in Figs.~1 and 4), and the {\em FeII}-classification even
clearer. Auroral [NII] 5755~\AA\ and [OI] 5577~\AA\ emission lines were
becoming visible, as well as nebular [OI] 6300, 6364~\AA. The intensity of
the auroral lines was comparable to that of the nebular transition lines,
indicating high densities in the outermost neutral external regions of the
ejecta where they formed.

On $\Delta t$=+18.4, $\Delta V$=2.6 mag, the [OI] lines grew in intensity more
than the other lines, reminiscent of the {\em [OI] flash} (cf. PG57) seen in other
novae at a similar evolution stage. The same happened to the [NII] lines and the
NII+OII+NIII blend at 4640~\AA\ on $\Delta t$=+36.4, $\Delta V$=3.3 mag when
FeII weakened and HeI turned into weak emission, corresponding to the {\em
[NII] flash} and {\em 4640-phase} observed in many novae at $\Delta
V$=3.0/3.5 mag (cf. PG57).

The reduced intensity of auroral with respect to nebular [OI] transitions in
May indicates a decline in electronic density in the outer ejecta, as
expected by the ongoing expansion and persistently low ionization stage.

\subsubsection{OI lines}

Emission lines from neutral oxygen offer several interesting applications
for novae, some of which have been discussed by Strittmatter et al. (1977),
Williams (1994), Bhatia \& Kastner (1995), Kastner \& Bhatia (1995), and
Osterbrock (1989). Formalism and atomic constants in the present section 
were adopted from these sources.

The [OI]  5577, 6300, 6364~\AA\ lines are among the first forbidden lines to
appear in nova spectra thanks to their high critical densities. They
persist even when [OIII] dominates the emission spectrum and high ionization
lines appear (see for example the simultaneous presence of [FeVII] and [OI]
lines in the spectra of Fig.~14). In the simplified case of spherically
symmetric, homogeneous ejecta, created by a steady wind and photoionized by
a central source, the [OI] lines trace ionization-bounded conditions and
are expected to form in the external, neutral regions. The ratio of the
6300, 6364~\AA\ nebular transitions under optically thin conditions is 3:1
from their transition probabilities, and it scales with optical depth $\tau$
(in the 6300~\AA\ line) as
\begin{equation}
\frac{F_{6300}}{F_{6364}} = \frac{1 - e^{-\tau}}{1 - e^{-\tau/3}}
\end{equation}
At typical oxygen abundance, ejected mass and outflow velocity for a nova,
[OI]~6300~\AA\ should turn optically thin within the first few days after
maximum brightness. Williams (1994) noted how the 6300/6364 \AA\ flux ratio
in novae is almost always lower than 3:1, indicating persistently optically
thick conditions in the lines. He proposed that optically thick [OI] lines
in novae come from small, very dense, neutral globules embedded in
ambient, ionized ejecta and that the globules are internally powered by
the decay of unstable isotopes produced by initial nuclear reactions (e.g.
Hernanz 2005, and reference therein, for a discussion of unstable nuclides).

Figure~1 illustrates the evolution of 6300/6364 \AA\ flux ratio over the
whole outburst of NCyg06. The lines first appeared on the April 12 spectrum,
$\sim$10 days after the outburst onset, with a 2.2:1 ratio, and then
initially evolved toward the 3:1 optically thin value. However, by late May
2006, when the decline of the nova slowed, the evolution of the 6300/6364
\AA\ flux ratio reversed its course and progressively decreased, reaching
the minimum value 1.1:1 by the time of second maximum. Such a low value
requires $\tau \approx$7. As soon as the nova entered the transition phase
immediately following the second maximum, with the ionization spreading
throughout the ejecta, the 6300/6364 \AA\ ratio rapidly converged to the 3:1
value expected for optically thin conditions. The 6300~\AA\ integrated line
flux, on the other hand, displayed only a modest decline during the
transition. Such a decline was confined to the phase when optically thin
dust condensed (cf. Sect. 7 below), and afterwards the 6300~\AA\ line 
progressively and rapidly regained the flux level it had reached 
{\em before} the transition phase.

  \begin{table}
     \caption{The optical depth, electron temperature and mass of neutral oxygen 
              in NCyg06 from the [OI] 5577, 6300 and 6364~\AA\ emission lines.
              The listed epochs are those of the spectra showing the 5577~\AA\ 
              line in emission.}
     \centering
     \includegraphics[width=8.8cm]{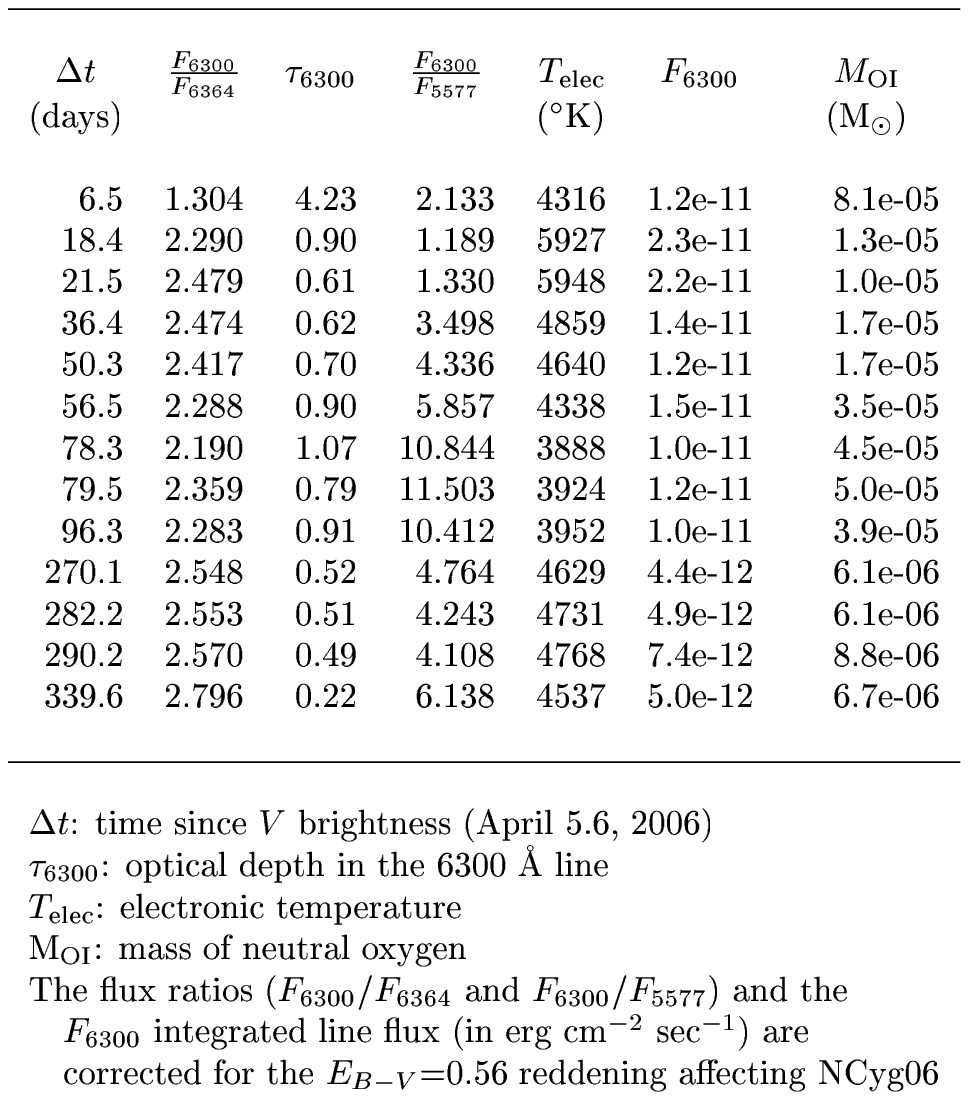}
     \label{tab6}
  \end{table}

The [OI] lines provide an estimate of the mass in the neutral region of the
ejecta.  Table~6 reports the results for some distinct epochs in the
evolution of NCyg06, where Eq. (4) is used to derive the optical depth from
the 6300/6364 \AA\ flux ratio. The high-density conditions in the neutral
regions make the auroral 5577 \AA\ line easily detectable and therefore
usable with the nebular 6300 \AA\ line to estimate the electron
temperature. How the latter correlates with the (reddening corrected)
5577/6300 \AA\ flux ratio for a wide range of 6300~\AA\ optical line depths
is shown in Fig.~7.  The proportionality between the integrated flux in the
6300~\AA\ line and the amount of neutral oxygen in the ejecta as a function of
optical depth and electron temperature is highlighted in Fig.~8 for the 7.2
kpc distance and $E_{B-V}$=0.56 reddening derived earlier for NCyg06. The
high density condition ruling the region of formation of OI lines is
highlighted by the OI $F_{8446}$/$F_{6300}$ emission line ratio. Its
reddening-corrected value was 16.5 on $\Delta t$=18.4 and $\Delta t$=21.5.
Comparing with the computation by Kastner and Bhatia (1995), it suggests a
density $\geq 10^{10}$~cm$^{-3}$. The ratio lowered to 0.4, 2.8, 0.4, and 
0.2 on $\Delta t$=186, 253, 340, and 527, respectively, indicating a 
density $\geq 10^{8}$~cm$^{-3}$.

  \setcounter{figure}{8}
  \begin{figure*}[!ht]
     \centering
     \includegraphics[width=18.0cm]{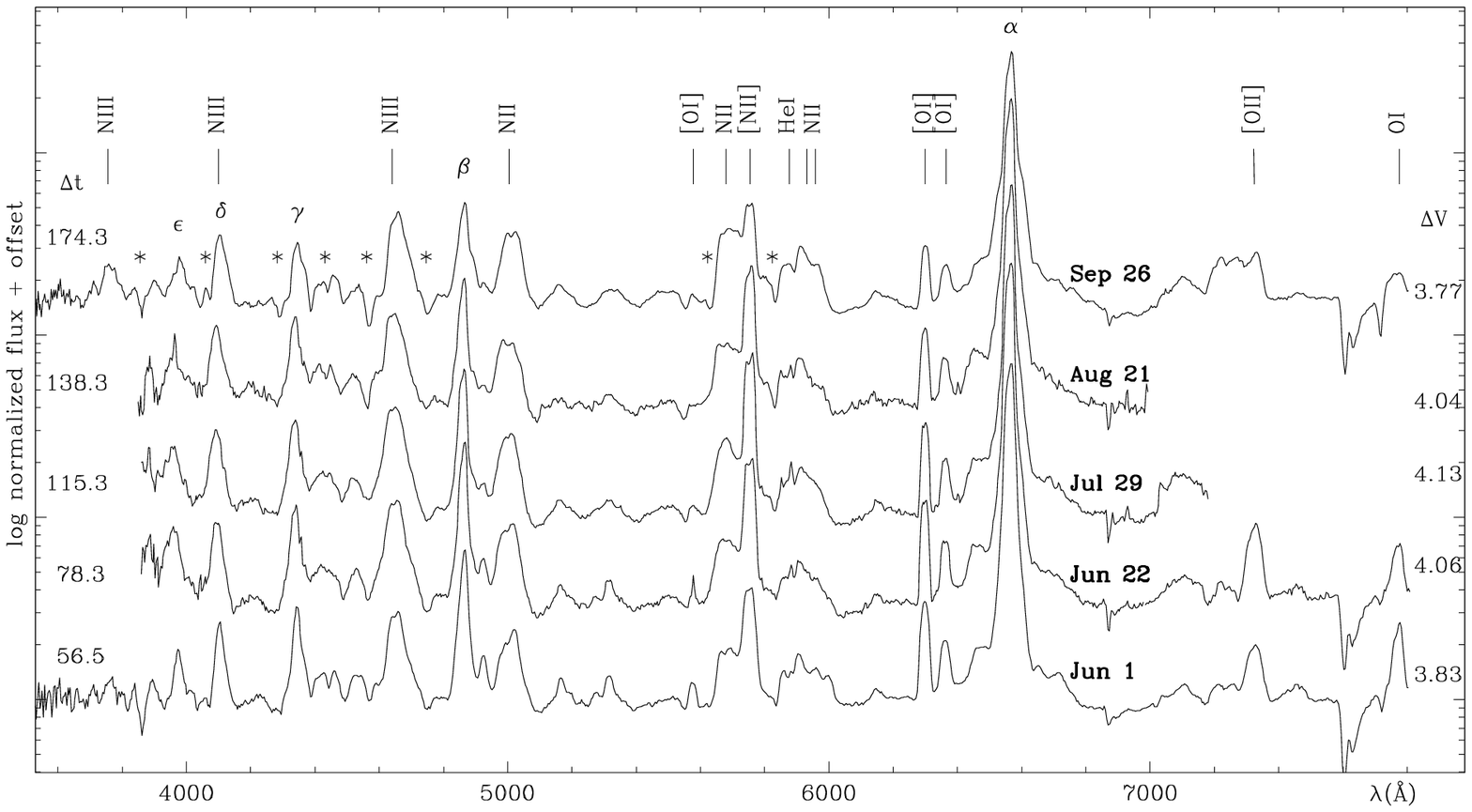}
     \caption{Spectroscopic evolution of Nova Cyg 2006 during the plateau
              phase in June-September 2006 (see Fig.~5 for plotting details). The 
              asterisks mark the positions of the principal Orion absorption 
              lines on the September 26 spectrum (see Sect. 6 for the line 
              list).}
     \label{fig9}
  \end{figure*}

The data in Table~6 suggest that, at the beginning of the nova evolution,
the neutral oxygen was present predominantly in the diffuse ejecta, while at
the later stages it only survived in isolated, dense, and neutral globules.
The mass of these globules accounts for $\sim$1/12 of the total mass of the
ejecta. This rough proportion can be derived by comparing the mass at first
[OI] detection ($\Delta t$=6.5 days), when nova ejecta were predominantly
neutral, with the mass at advanced decline ($\Delta t$$>$260 days) when the
nova ejecta were essentially completely ionized (density bounded).

It will be shown in Sect.~11 that the ejecta of NCyg06 are oxygen enhanced
(an O/H ratio $\sim$32$\times$ larger than in the Sun). On $\Delta t$=6.5
days, all oxygen in the NCyg06 ejecta was neutral, and therefore the mass
derived for neutral oxygen corresponds to the total mass of oxygen in the
ejecta ($M_{\rm O}$=$M_{\rm OI}$). Noting that the oxygen mass fraction
derived in Table~9 from nebular analysis is 0.163 and that dust had not yet
formed during early decline, the total mass of the ejecta can be estimated
as
\begin{equation}
M_{\rm ejecta} = \frac{M_{\rm O}}{0.163} \approx 5 \times 10^{-4} ~~~~{\rm M}_\odot
\end{equation}
where $M_{\rm OI}$=8.1$\times$10$^{-5}$~M$_\odot$ is taken from Table~6.
An independent estimate for $M_{\rm ejecta}$ will be obtained in Sect.~11 
from analysis of the emission line spectrum in the nebular phase.

It is also noteworthy in Table~6 that the amount of neutral oxygen parallels
the photometric evolution. In fact, in Sect.~5.1 we discussed how by
June 1, 2006 the nova was departing from a pure exponential decline (to
progressively slow its decline and later rise to a second maximum). At the
same time, the amount of neutral oxygen (and presumably of neutral ejecta)
was increasing, as if the ionization boundary was retracting (in a mass
sense) through the ejecta.

The intensity of the OI 8446~\AA\ emission line under normal recombination,
optically thin conditions should be 3/5 of the OI 7774 line. The reddening-corrected 
flux ratio observed in NCyg06 is $F_{8446}$/$F_{7774}$=7.7 on
$\Delta t$=18.4 (decline from first maximum), 21.5 on $\Delta t$=290
(decline from second maximum), and 12.5 on $\Delta t$=340 (nebular condition
during the advanced stages of the final decline). The inversion in
intensity between the two OI lines is usually associated with fluorescence
pumped by absorption of hydrogen Lyman-$\beta$ photons, as first pointed out
by Bowen (1947). For the Lyman-$\beta$ fluorescence to be effective, the
optical depth in H$\alpha$ should be large, presumably owing to the
population of the $n=2$ level by trapped Lyman-$\alpha$ photons. The
$F_{8446}$/$F_{H\alpha}$ under optically thin, low ionization conditions and
typical nova chemical abundances is quite low, $\sim$10$^{-3}$ (Strittmatter
et al. 1977). The evolution of the $F_{8446}$/$F_{H\alpha}$ flux ratio for
NCyg06 is presented in Fig.~1. It reached a much higher $\sim$0.5 value
during early evolution (on $\Delta t$=18.4 and 21.5). It indicates a very
large optical depth in H$\alpha$, in agreement with the ejecta
still being predominantly neutral and very dense at that time.  At later
times the picture was more confused due to the clumpiness of OI-emitting
regions described earlier and the growing [NII] contribution to the
H$\alpha$ observed flux. Nevertheless, even at advanced evolutionary stages,
the $F_{8446}$/$F_{H\alpha}$ did not decline below 0.05, indicating
persistent (even if reduced) optically thick conditions in Halpha. This
agrees with the evidence from the $F_{8446}$/$F_{7774}$ ratio that
Lyman-$\beta$ fluorescence has remained effective throughout all NCyg06
evolutionary stages.

\section{The plateau phase}

When NCyg06 left the exponential decline of Eq.(1) around June 1, it had
magnitudes and colors of $V$=11.83, $B-V$=+0.28, $V-R_{\rm C}$=+1.44,
$R_{\rm C}-I_{\rm C}$=+0.20 and $V-I_{\rm C}$=+1.64. A minimum brightness
was reached on July 15 at $V$=12.15, $B-V$=+0.15, $V-R_{\rm C}$=+1.28,
$R_{\rm C}-I_{\rm C}$=$-$0.01 and $V-I_{\rm C}$=+1.27. During the following
rise in brightness, the same $V$-band brightness as for June~1 was regained
111 days later on September 20, when it was $V$=11.83, $B-V$=+0.16,
$V-R_{\rm C}$=+0.91, $R_{\rm C}-I_{\rm C}$=+0.07 and $V-I_{\rm C}$=+0.99. 

The significantly bluer colors on September 20 compared to June 1 indicate
that the nova had appreciably evolved. This is clear from the evolution of
the spectrum (illustrated in Fig.~9) during this plateau period.
Progressively, FeII emission disappeared, while NII/H$\beta$ and
NIII+H$\delta$/H$\gamma$ increased, and [NII] decreased with respect to NII and
increased compared to [OI]. However, the most interesting feature of this
plateau period is the re-emergence of the Orion absorption system. The Orion
absorptions were last seen on the May 11 spectrum and marginally still
visible on June 1.  On the August 21 and September 26 spectra of Fig.~9, the
most notable Orion absorption lines are those of HeI 3889 \AA\ ($-$2140 km/s
radial velocity), 4471 ($-$2065 km/s), 5876 ($-$2260 km/s), OII 4321
($-$2125 km/s), NII 4601 ($-$2100 km/s), 4788 ($-$2150 km/s), 5672 ($-$2100
km/s). Particularly interesting is the re-emergence of NIII 4097-4103
($-$1950 km/s) on the September 26 spectrum (compare the $H\delta$ blue wing
with that of the June 1 spectrum on Fig.~9). The average velocity of the
Orion absorptions on September 26 was $-$2100~km/s, significantly higher
than on June~1 when the Orion spectrum was last seen (cf. Table~5). The
acceleration seen in the Orion absorptions had the reverse sign for NIII
4097-4103: $-$2200 on June 1 and $-$1950 km/s on September 26.

During this plateau period, we obtained high-resolution H$\alpha$ profiles
on June 1 and 24, July 18 and 29, August 31 and September 26. The H$\alpha$
profles from June 1 to August 31 are all the same and very similar to that
of June 24 shown in Fig.~12, which is characterized by a multi-peaked top
and by 1340~km/s full width at half intensity. On the other hand, the
September 26 profile is characterized by the emergence of a hazy, broad and
featureless pedestal whose full width at zero intensity approaches 6000
km/s, while the width at half intensity is $\sim$4200~km/s (matching the
$\sim$$-$2100~km/s expansion velocity of the Orion absorption lines). This
pedestal grew conspicuously in relative intensity by the time of secondary
maximum (cf Fig.~12). It is worth noticing that the decline of the Orion
absorptions in novae is usually accompanied by the emergence in emission
line profiles of hazy and broad components, even if Balmer lines do not
always display them.

\section{The second maximum}

The second maximum was reached by NCyg06 at $V$=9.95 around HJD 2454071.4
(December 1, 2006) at $\Delta t$=+239.0 and $\Delta V$=1.95, with colors
$U-B$=$-$0.29, $B-V$=+0.28, $V-R_{\rm C}$=+0.68, $R_{\rm C}-I_{\rm
C}$=+0.28, $V-I_{\rm C}$=+0.98. The emission lines at that time had a
minimal impact on the overall energy distribution, and the intrinsic
(B-V)$_\circ$=$-$0.28 corresponds to that of a B0 giant, compared to an F0
giant at first maximum. Compared to the first maximum, the colors at the
second maximum were bluer by the following amounts: $\Delta
(B-V)$$\approx$$-$0.60, $\Delta (V-R_{\rm C})$$\approx$$-$0.42, $\Delta
(R_{\rm C}-I_{\rm C})$$\approx$$-$0.20 and $\Delta (V-I_{\rm
C})$$\approx$$-$0.52 (colors at $t$=0 are extrapolated by parabolic fitting
to the early decline data in Table~1 and Fig.~1).  The time required to rise
the last magnitude in the $V$ band was 22 days, to decline the first two
magnitudes was 9.6 days, and to decline the first three magnitudes was 11.9
days.  The brightness and color evolution around the second maximum is
highlighted in Fig.~10.

The overall spectroscopic aspect of NCyg06 at the second maximum was
appreciably different than at the first one, as seen when comparing the spectral evolution
in Fig.~5 with Fig.~11). In addition to the Balmer series, only low-excitation 
absorption lines (mainly CaII, NaI, FeII) were seen at the first
maximum, in agreement with the F0 underlying spectral energy distribution.
At second maximum, the underlying continuum was much hotter and the
excitation of the absorption spectrum correspondingly much higher. The
strongest, non-Balmer absorption lines were: CII 6789, 6746, 5143; NII 6482,
5678, 5667, 5045, 5007, 4791, 4602; NIII 4639, 4513, 4100; OII 4495, 4468,
4416, 4318, 4070 and HeI 6678, 5876, 4471, 3889. Their mean radial velocity
was $-$2190~km/s for CII, $-$2135~km/s for NII, $-$2070~km/s for OII, and
$-$2170~km/s for HeI.  The absorption spectrum at second maximum represents
the re-emergence of the Orion system that disappeared during the plateau
phase.

Quite interesting are the absorption components seen in H$\alpha$ during the
second maximum. The spectrum for November 5, 2006 in Fig.~12 is
representative of them. The underlying emission is obviously much more
structured compared to the simple, symmetric one seen during first maximum
(cf. Fig.~6), and therefore unambiguous identification of weaker absorptions
is severely affected. Nevertheless, at least two strong absorptions are
safely identified: one centered at $-$905~km/s (VHM=145~km~sec$^{-1}$,
e.w.=1.90~\AA, integrated flux=8.4$\times 10^{-13}$ erg cm$^{-2}$
sec$^{-1}$), the other at $-$2215~km/s (VHM=355~km~sec$^{-1}$,
e.w.=7.67~\AA, integrated flux=2.0$\times 10^{-12}$ erg cm$^{-2}$
sec$^{-1}$). They could be tentatively identified with the re-emergence of
{\em principal} and {\em diffuse-enhanced} absorption systems seen during
the decline from the first maximum (cf. Table~4). On November 22, the faster
of the two components had grown to an integrated flux of 3.5$\times
10^{-13}$ erg cm$^{-2}$ sec$^{-1}$ and an equivalent width 9.8~\AA.

Similar to what has just been described for the absorption spectrum, the
emission spectrum at second maximum was quite different from that at first
maximum, reflecting the much higher temperature of the underlying continuum.
At second maximum, FeII emission lines were not seen and were replaced by
NII, NIII, OII, [OI], and HeI. [NII] 5755 disappeared, and the
$F_{6300}$/$F_{6364}$ ratio approached 1.1:1 (cf. Fig.~1 and Sect.~5.3.3).
Overall, the strong decrease in equivalent width of the emission lines at
secondary maximum (cf. H$\alpha$ behavior in Fig.~1) essentially stemmed from
the increase in the underlying continuum. The evolution of integrated flux
of emission lines went unaffected through the development of second maximum,
as the evolution of [OI] and H$\alpha$ in Fig.~1 illustrates.

  \begin{figure}
     \centering
     \includegraphics[width=8.8cm]{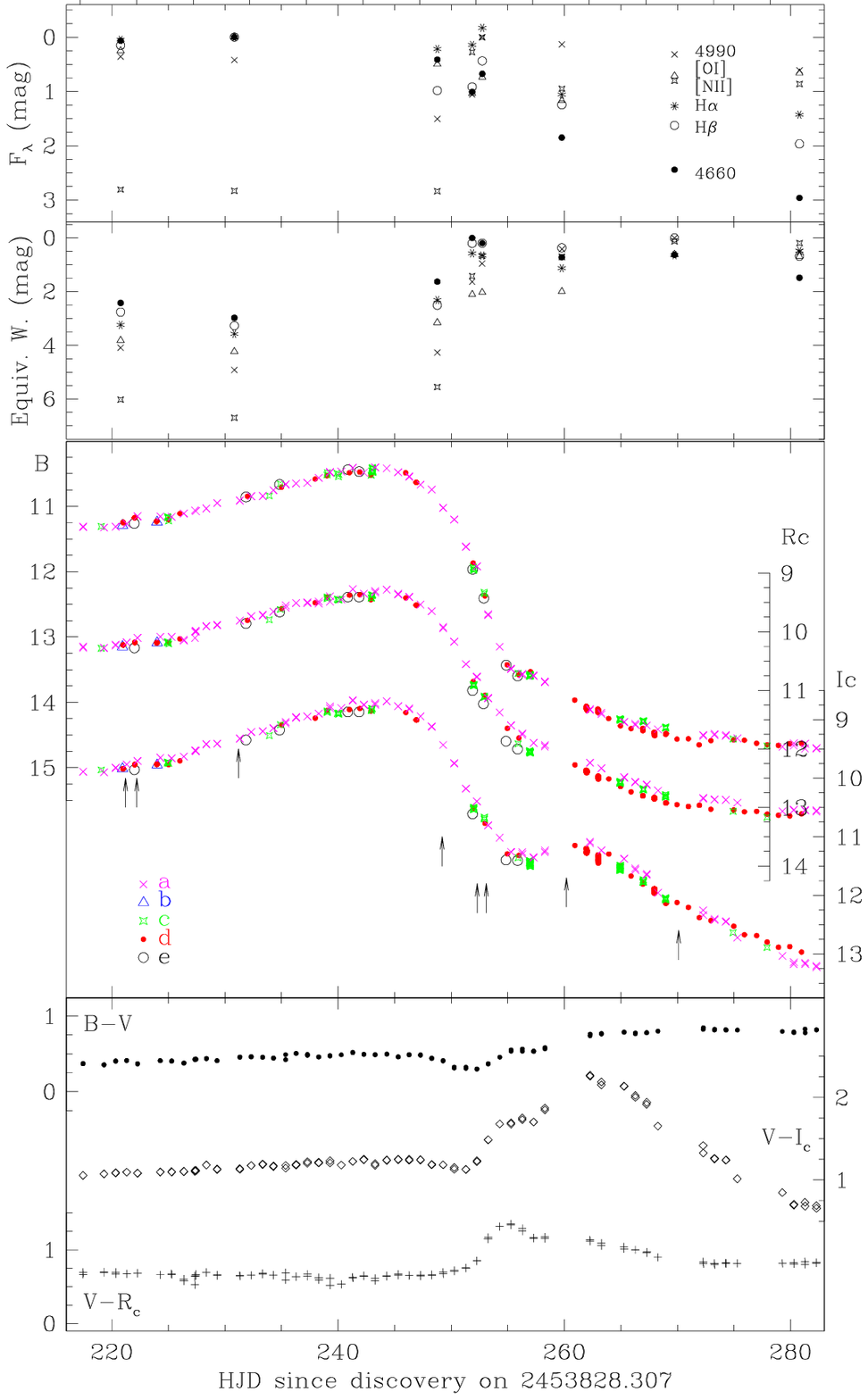}
     \caption{{\em Middle panel}: Photometric evolution of Nova Cyg 2006
              around second maximum. $B$-band lightcurve is at the top,
              $R_{\rm C}$ center, $I_{\rm C}$ bottom. For telescope
              identification letters see sect. 2.1. {\em Bottom panel}:
              color evolution as recorded with telescope $a$. 
              {\em Upper panels}: evolution of integrated
              flux and equivalent width of selected emission lines,
              expressed in magnitudes with respect to highest value. The
              {\em 4660} blend is initially dominated by NIII, than HeII
              become a relevant contributor.  The {\em 4990} blend is 
              initially due to NII, then [OIII] becomes the dominant contributor.}
     \label{fig10}
  \end{figure}

\section{Final decline}

Soon after the second maximum was reached on December 1, NCyg06 entered the final
decline that is continuing up to now. During this phase, NCyg06 resumed the
normal spectroscopic evolution of a nova that was interrupted by the plateau
phase and rose to secondary maximum. The decline is characterized by a sharp
change in the slope of the lightcurve occurring around December 12, as
illustrated by both Figs.~1 and 10. That date also marks the
transition of NCyg06 from stellar to nebular spectral type. 

\subsection{Decline prior to December 12}

Between the second maximum and December 12, the nova became fainter by
$\Delta V$=2.8~mag in 11 days, with $t_{2}^{V}$=9.6$\pm$0.2 and 
$t_{3}^{V}$=11.9$\pm$0.3~days. The slope of the declines from the first and 
second maxima are quite different, as the respective $t_{2}$/$t_{3}$ suggest and
a comparison of Figs.~4 and 10 supports. The rate of decline from first maximum 
was slowing with time, that from second one was increasing with time.
The $B-V$ color initially turned bluer as the stellar
continuum turned hotter, but this was short-lived and by December 8 it
reversed the trend. By December 13 it had resumed the smooth course it has been
displaying since the earliest outburst phases (cf. Fig.~10). The $V-R_{\rm C}$
color increased almost exponentially from December 7 to 12, as a result of
the rapidly growing relative contribution of H$\alpha$ to $R_{\rm C}$-band
(cf. H$\alpha$ equivalent width evolution in Fig.~1). The extra flux in the
$I_{\rm C}$-band, clearly influencing $R_{\rm C}-I_{\rm C}$ and $V-I_{\rm
C}$, will be discussed in Sect.~9 below.

The strong multiple absorptions seen in the Balmer lines and the Orion
absorption system, so prominent at the time of second maximum, quickly
vanished after the maximum.  The only absorption lines still present in the
December~10 spectrum are NII 4601, NIII~4097-4103, OIII 5592~\AA, and a
weak, residual $-$1750~km/s absorption component in H$\alpha$. They too had
vanished by December 12. The emission line spectrum during decline from
second maximum changed much less than during decline from primary maximum.
The ionization/excitation increased only modestly, while there was a strong increase in
equivalent width of the emission lines. The major change was 
seen in the optical depth of [OI] that soon started to decline from
$\tau_{6300}$$\approx$7 at maximum, to $\tau_{6300}$$\approx$0.15 on
December 12.

  \begin{figure*} 
    \centering 
     \includegraphics[width=18.0cm]{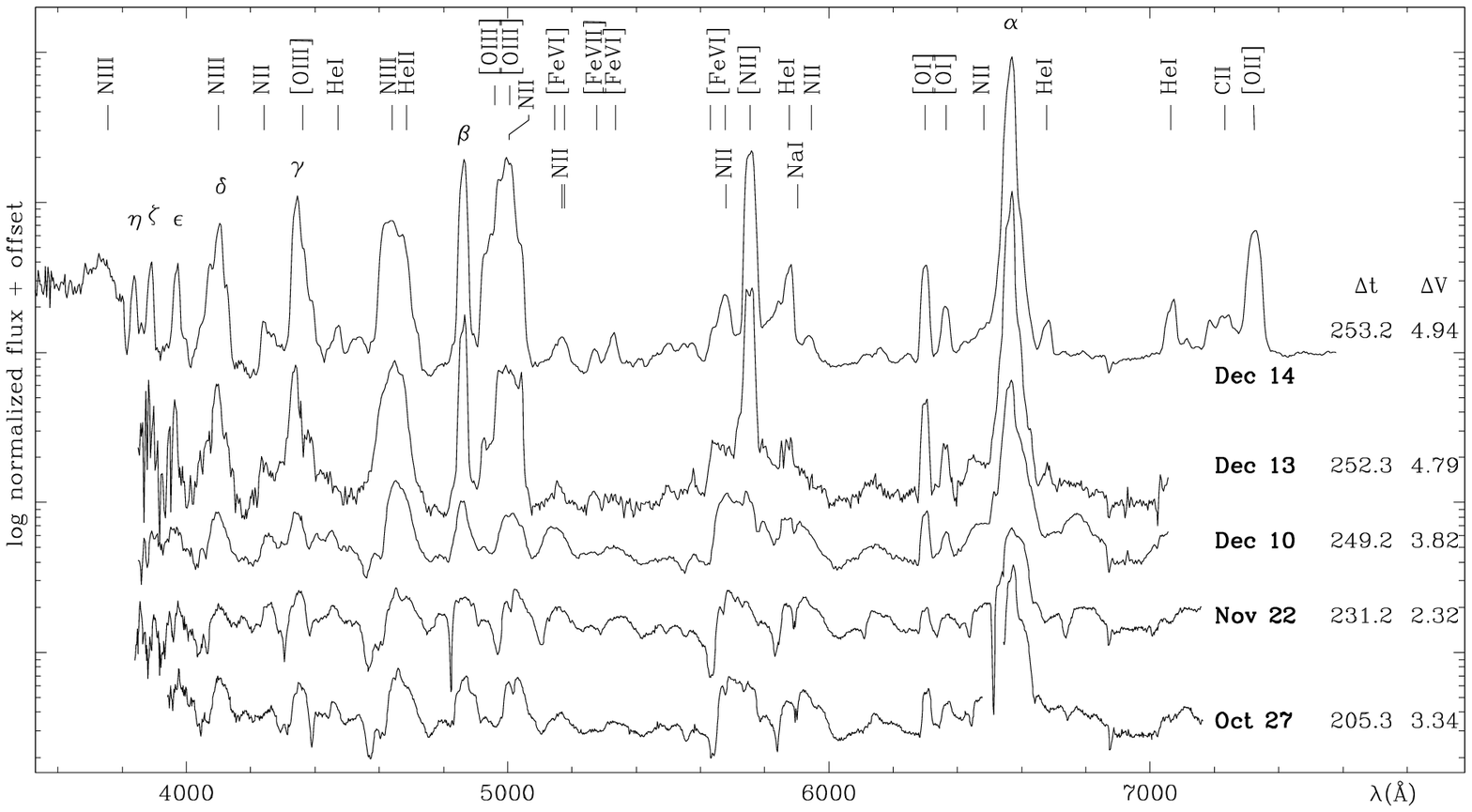}
     \caption{Spectroscopic evolution of Nova Cyg 2006 during second maximum 
              in autumn 2006 (see Fig.~5 for plotting details).}
     \label{fig11} 
     \end{figure*}

\subsection{Decline after December 12}

Around December 12 the rapid decline of NCyg06 from the second maximum
suddenly slowed down, settling to a $<$$\Delta B$$>$=0.0088~mag~day$^{-1}$
straight linear slope over the range +300$\leq$$\Delta t$$\leq$+460. This
transition in the photometric evolution corresponds to the transition from
stellar to nebular spectral types. In fact, the spectrum for December~10 in
Fig.~11 is still dominated by the continuum emission with a modest
contribution from emission lines. Three days later (cf. December 13 spectrum
in Fig.~11), the spectrum had already turned into the nebular type, with a
huge increase in the equivalent width of emission lines and weakening of the
underlying continuum. In particular, [OIII] 4363 was emerging on the red
wing of the H$\gamma$ profile, [OIII] 4959, 5007 were already key contributors
to the strong emission blend with NII centered at 4990~\AA, [NII]~5755 was
already stronger than H$\beta$, and HeII 4686 was becoming detectable
superimposed over the huge NIII emission at 4640~\AA. During the first weeks
of the nebular state, nebular lines appeared double-peaked with a
peak-to-peak velocity separation of 595~km/s for [OI]~6364, 815~km/s for
[NII]~5750, and 950~km/s for [OIII]~5007~\AA.

With the transition from stellar to nebular type, the emission from the
rapidly growing fraction of ionized ejecta compensated for the rapid photometric
decline associated with the retracting pseudo-photosphere, which was
increasing in temperature and therefore was shifting the peak of its emissivity 
to shorter and shorter UV wavelengths. The reprocessing into
$U$$B$$V$$R_{\rm C}$$I_{\rm C}$ bands and longer wavelengths of ultraviolet
flux absorbed by the ionized ejecta resulted in the leisurely $<$$\Delta
B$$>$=0.0088~mag~day$^{-1}$ decline rate and in the protracted visibility of
the nova for such a long period. Without reprocessing ejecta, the central
source would have become unobservable in the optical by January 2007.

  \begin{figure}
     \centering
     \includegraphics[width=8.0cm]{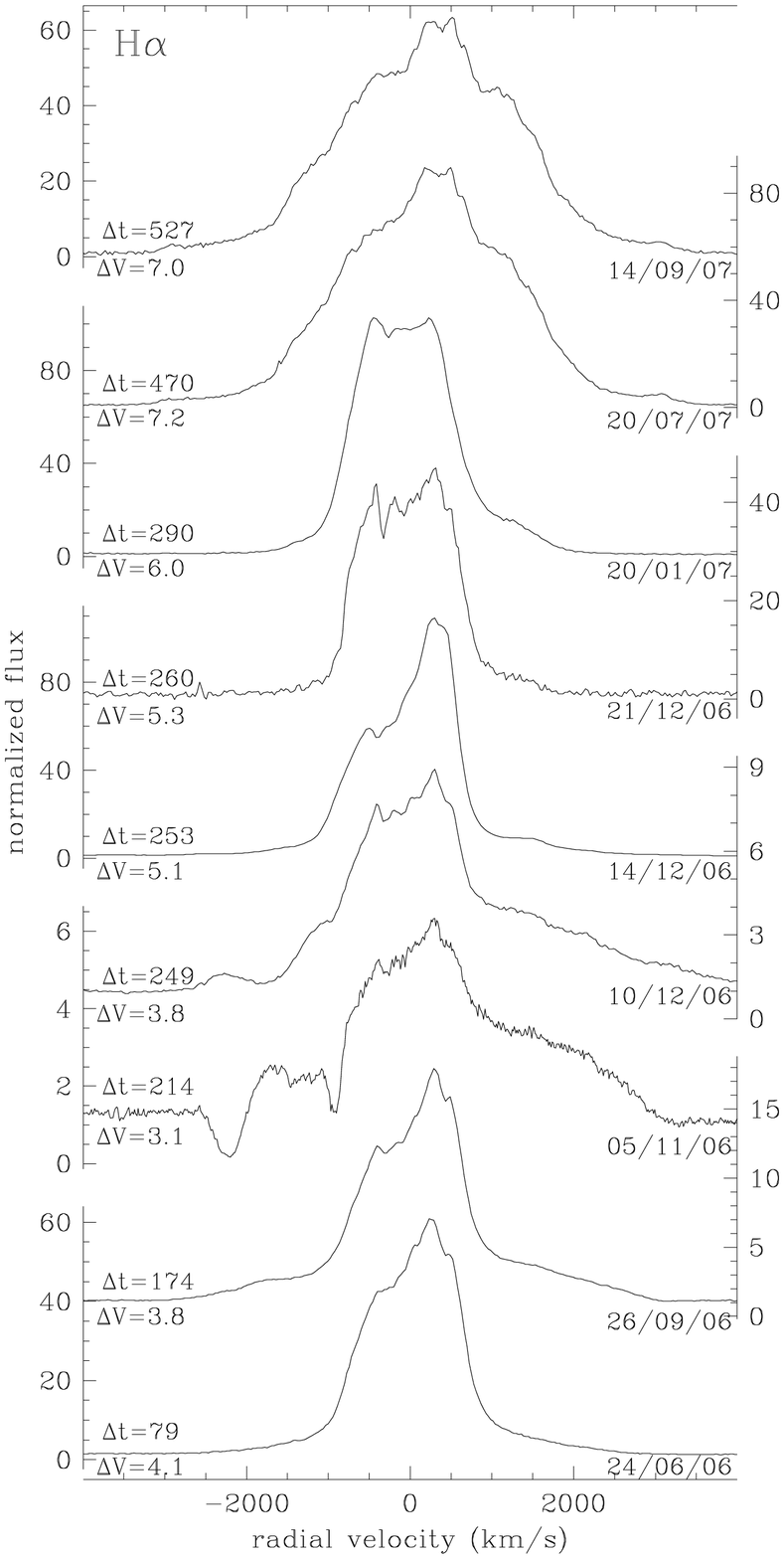}
     \caption{Evolution of H$\alpha$ profile during the plateau phase, the
              second maximum and the advanced decline. $\Delta t$ and
              $\Delta V$ are days and magnitudes from maximum.}
     \label{fig12}
  \end{figure}

\subsection{The fast and tenuous wind of mid 2007}

As illustrated by the spectra in Fig.~14, in July-September 2007 the
emission lines grew in width. The FWHM of [OI] lines increased from
1325~km/sec on March 11 spectra, to 1870~km/sec on July 20, and 1890~km/sec
on September 19, and [NII] lines from 1570~km/sec to 2070 and 2135~km/sec,
respectively. This trend is illustrated in Fig.~1, which also shows how
the photometric decline almost stopped during the same period, with an
extra $\sim$0.6~mag flux in the $B$ and $V$ bands with respect to
extrapolated earlier lightcurves. We interpret these events as due to the
emergence of a fast, tenuous wind from the central star. Its effects are
equally evident in the evolution of the blended H$\alpha$+[NII] profile
in Fig.~12.

\section{Dust formation}

Infrared photometry of NCyg06 is available from the literature for only three
dates: April 30, June 14, and December 12, 2006 (Russell et al. 2006; Mazuk
et al. 2006; Lynch et al. 2006). These IR magnitudes were combined with our
simultaneous $U$$B$$V$$R_{\rm C}$$I_{\rm C}$ from Table~1 to construct the
spectral energy distributions shown in Fig.~13, where the interpolating
blackbodies have been reddened to the $E_{B-V}$=0.56 affecting NCyg06. On
April 30 and June 14, the $U$$B$$V$$R_{\rm C}$$I_{\rm C}$$J$$H$$K$ spectral
energy distribution can be ascribed entirely to emission from the
pseudo-photosphere with no contribution from warm dust, while on
December~12 ($\Delta t$=+251$^d$) thermal emission from dust is quite
apparent.

The dust detected on December 12 was optically thin. The energy re-emitted
by dust in the infrared is energy absorbed from energy radiated by the nova,
and therefore the dust optical depth is $\tau$=$L_{\rm IR}$/$L$. $L$ is
difficult to estimate directly at the time of dust condensation because most
of it is radiated in the ultraviolet as indicated by the nebular and high
ionization condition of the ejecta and illustrated by the spectral energy
distribution of Fig.~13. However, theoretical models argue for a prolonged
{\em plateau} period of constant luminosity experienced by novae following
optical maximum (Prialnik 1990, and references therein), which is confirmed
by multi-wavelength observations of well studied novae (Gallagher 1977;
Gallagher and Starrfield 1978; Gehrz et al. 1998).  It can therefore be
written that
\begin{equation}
\tau_{\rm dust} = \frac{L_{\rm IR}}{L_{\rm plateau}}
\end{equation}
In Sect.~11 below, we show from photo-ionization modeling that on
January 20, 2007 ($\Delta t$=+290) the bolometric luminosity of NCyg06 was
$M_{\rm bol} = -6.45$ corresponding to $L_{\rm bol}$= 1.3$\times$10$^{38}$
erg~sec$^{-1}$. We assume this was valid for December 12, 2006 as well.
The luminosity radiated by the 1550~K blackbody fitting the December 12 dust
emission over the $L$, $M$, and $N$-band photometric points of Fig.13 is
$L_{\rm max}$= 2.6$\times$10$^{37}$ erg~sec$^{-1}$. The optical depth of the
dust for December~12 is therefore $\tau_{\rm dust}$$\sim$0.2, indicating
optically thin conditions.

The infrared data available for NCyg06 do not enable a derivation of physical
properties of the dust. However, for sake of discussion, we assume that the
dust grains condensed in NCyg06 followed the mean properties observed in
other novae (Gehrz 1988; Mason et al. 1996; Evans et al. 1997; Gehrz et al.
1998), i.e. they were small carbon grains (radius
$a$$\leq$1~$\mu$m, density $\rho$$\sim$2.3~gr~cm$^{-3}$), for which the
Planck mean emission cross section goes as $Q_e = 0.01 a T^{2}_{dust}$. Under
these assumptions, the mass of the dust in NCyg06 on December 12 is
\begin{equation}
M_{dust} = 1.17 \times 10^6 \rho T^{-6}_{dust} \left( 
\frac{L_{\rm IR}}{L_\odot} \right) = 1.3\times 10^{-9}~~~{\rm M}_\odot 
\end{equation}
which is quite small. Comparing with the amount of gas in the ejecta
estimated below in Eq.(9), the gas-to-dust ratio in the ejecta on that date
is $\left( \frac{M_{\rm gas}}{M_{\rm dust}} \right) \approx 3 \times 10^5$.
A similarly high value of the gas-to-dust ratio has been previously observed
in Nova Vul 1984~N.2, which, like NCyg06, displayed a large oxygen
overabundance ([O/H]=+1.6, Gehrz et al. 1998). It should also be noted that
dust did not appear to form in some other novae, adequately observed in the
infrared.

As a first guess, it may seem that the hard radiation field from the central
star, responsible for the ionization of the ejecta, should suppress dust
grain formation (e.g. Gallagher 1997). Actually, observations of well-studied 
novae indicate that dust condenses when the ejecta become optically
thin in the near ultraviolet and the nova spectrum shifts from stellar to
nebular type. Shore \& Gehrz (2004) proposed that ionization may be a
promoting agent of dust condensation via ionization-mediated kinematic
agglomeration of atoms onto molecules and small grains through induced
dipole interactions. It is interesting to note that dust was detected in
NCyg06 on December 12, right when the transition was taking place. When comparing
the December 10 and December 13 spectra in Fig.~11, the huge increase in both
the amount and excitation degree of nebular emission is outstanding.
Description of the infrared observations by Rayner et al. (2006) indicates
that dust emission was absent on November 30 (when the nova was peaking at
second maximum), and was still present on December 20 when the
transition to a nebular spectrum was completed ([OIII] 4949 and [NII]
5750~\AA\ appreciably stronger than H$\beta$). They estimated a black-body
temperature of 1410~K beyond 2~$\mu$m. Once corrected for the $E_{B-V}$=0.56
reddening affecting the nova, their estimate is in fair agreement with our
derived 1550~K for December 12 data.

Such a temperature is unusually high for dust grains in novae (cf. Gehrz
2002). However, it has to be noted that on December 12 the dust had just
begun to form in NCyg06 (a matter of a few days or even hours), it had a
quite thin optical depth and it was quite low in mass. These are conditions
away from the average ones applying to the novae for which dust properties
have been studied in detail. We do not discuss the dust temperature in
NCyg06 further, since we lack the necessary infrared data.  In particular,
contribution by emission silicate features at 10~$\mu$m and hydrocarbon
features at 3.3, 3.4, and 11.3~$\mu$m have been observed in some novae
superimposed onto the carbon dust continuum during the optically thin phase
(for ex. in V705~Cas by Gehrz et al. 1995, or V842~Cen by Gehrz et al.
1990). The presence of these (strong) emission features may alter the slope
of the spectral energy distribution and consequently spoil the significance
of temperature estimates from black-body fitting to broad-band photometric
measurements.

  \begin{figure}
     \centering
     \includegraphics[width=8.0cm]{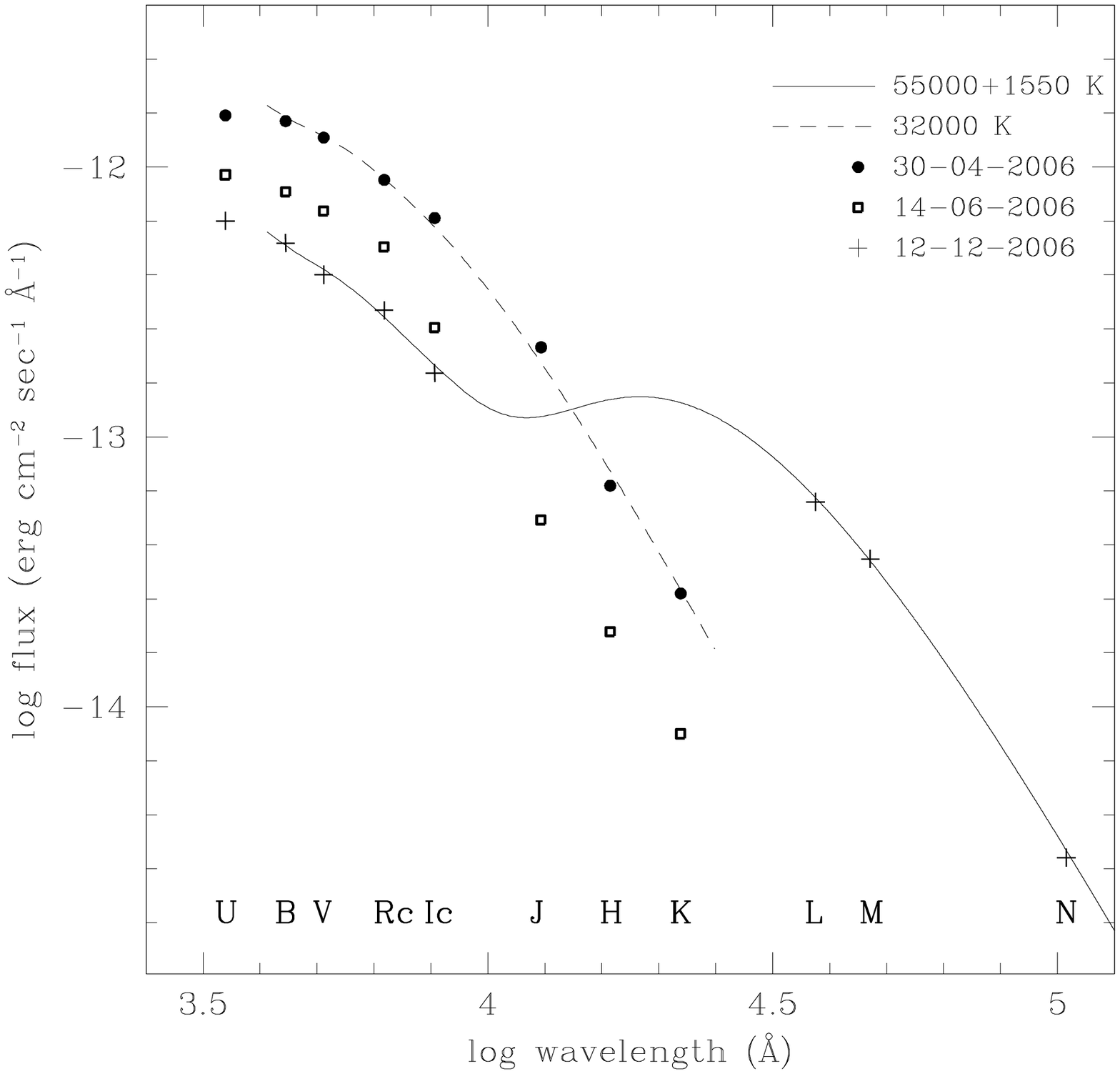}
     \caption{The optical and infrared spectral energy distribution of Nova Cyg
              2006 showing the presence of optically thin dust emission on
              December 2006. Optical data are from Table~1, and infrared data from
              Russell et al. (2006), Mazuk et al. (2006), and Lynch et al.
              (2006). The curves are fits with black bodies at the indicated
              temperatures and reddened by $E_{B-V}$=0.56 following a
              $R_V$=3.1 standard law.}
     \label{fig13}
  \end{figure}

An intriguing feature of NCyg06 is the $I_{\rm C}$ band excess it displayed
during the second half of December (see Fig.~10), soon after the dust
detection on December 12.  Following the second maximum, all bands declined
steeply and in a similar fashion, with a marked slope change around
December~12 when the transition from stellar to nebular spectrum took place
and the ionization spread widely through the ejecta. From that date, all bands
declined linearly with time, except $I_{\rm C}$, which displayed a pronounced
extra-emission of $\Delta I_{\rm C}$$\approx$1 mag amplitude peaking around
December 21 (see Fig.~10). During this time interval, we did not obtain
spectra covering the $I_{\rm C}$ wavelength range. The closest spectra
were taken on December~12, when the transition to nebular spectrum was
already rapidly  proceeding, and it shows only a modest OI~8446~\AA\
emission line and a weak Paschen continuum in emission, neither of which
are able to account for a sizable fraction of the $I_{\rm C}$ flux. We are
therefore inclined to exclude a flaring of emission lines selectively
concentrated over its wavelength range as the cause of extra emission in the
$I_{\rm C}$ band during the second half of December. This is also supported
by the evolution of the $V-R_{\rm C}$ color illustrated in Fig.~10, which was
dominated by the huge increase in the  H$\alpha$ equivalent width (cf. Fig.~1)
induced by the nova transition to the nebular stage (on the December 21
spectra, at the time of $I_{\rm C}$ band bump maximum, H$\alpha$ was
accounting for 0.63 mag of the whole flux in the $R_{\rm C}$ band). The
$V-R_{\rm C}$ color evolution is smooth and linear throughout the $I_{\rm
C}$ bump development.

The only feasible source of extra-emission that could have been responsible
for the $I_{\rm C}$ December bump seems to be the dust. For dust emission to
have leaked into the $I_{\rm C}$ band, it must have condensed into larger
quantities and/or peaked to hotter temperatures than on December~12. A
choice between the two alternatives could be possible only with the support
of (missing) infrared data covering the second half of December.

\section{Comparison with V1493 Nova Aql 1999a}

The lightcurve of NCyg06 has been highly peculiar, but not completely unprecedented.
A close match is represented by Nova Aql 1999 N.1 (=V1493 Aql), and Fig.~15
compares the $V$-band lightcurves of the two novae. The similarity is
striking, with V1493 Aql requiring 5.0$\times$ less time to go from
primary to secondary maximum, even if its $t_2$$\approx$7 and
$t_3$$\approx$23 days (Venturini et al. 2004) are not that different from
$t_2$=10.4 and $t_3$=24 of NCyg06. Both novae seem to have orbital periods
above the period gaps of cataclysmic variables, 5.0 hours for NCyg06
(Goranskij et al. 2006) and 3.7 hours for V1493 Aql (Dobrotka et al. 2006).
Both novae belong to the FeII class and shared similar expansion velocities
at first (FWHM=1700 km/s for V1493 Aql, Arkhipova et al. 2002) and second maximum
(FWHM=2000 km/s for V1493 Aql, Venturini et al. 2004). The spectral evolution of
V1493 Aql has been poorly monitored. However, what is recorded  suggests a path
parallel to that of NCyg06: ($a$) low ionization conditions at both primary
(only Balmer and FeII lines in emission in optical spectra, Tomov et al.
1999) and secondary maxima (only Paschen, Bracket, OI and NI lines in
emission with marginal HeI in IR spectra, Venturini et al. 2004), ($b$)
slight increase in ionization between the two maxima with appearance of OII,
[OII], NII and [NII] emission lines and disappearance of FeII (Arkhipova et
al. 2002), ($c$) spectra obtained during the advanced decline characterized
by nebular conditions, with [OIII], [NII] and NIII among the strongest emission
lines (Arkhipova et al. 2002). 

  \begin{figure*} 
    \centering 
     \includegraphics[width=18.0cm]{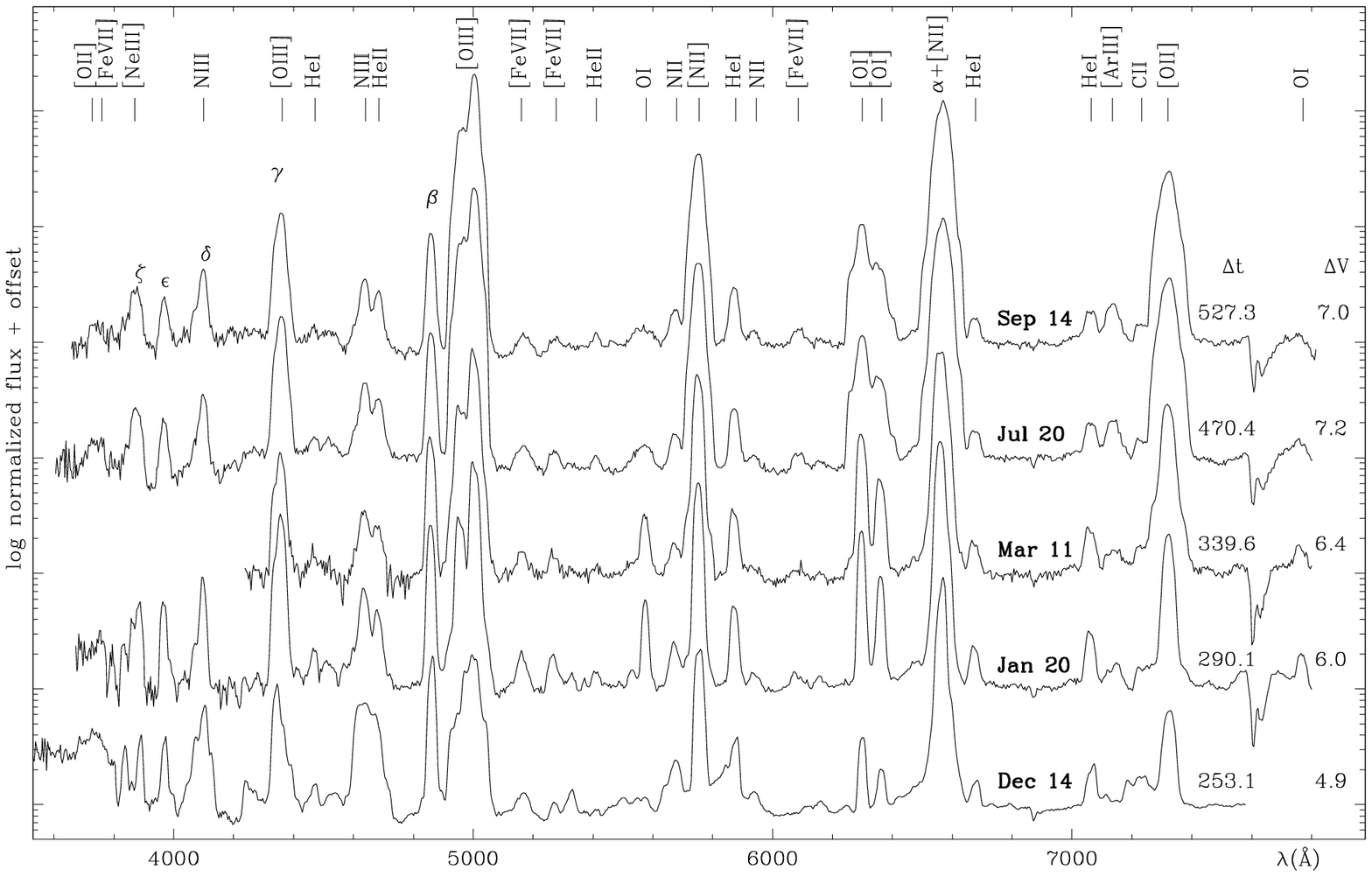}
     \caption{Spectroscopic evolution of Nova Cyg 2006 during the advanced decline
              from December 2006 to September 2007 (see Fig.~5 for plotting   
              details).}
     \label{fig14} 
     \end{figure*}

The only possible major difference could be in the color evolution.
Bonifacio et al. (2000) assembled heterogeneous $B-V$ measurements published
in several IAUC with some of their own, and concluded that V1493 Aql was
hotter at maximum light ($B-V$$<$0.4) than at secondary peak
($B-V$$\sim$1.2) or plateau phase ($B-V$$\sim$0.9), with a bluer color
during advanced decline ($B-V$$\sim$0.7). Unfortunately, the intrinsic
variability of the comparison stars chosen by Bonifacio et al. to reduce
their own measurements and the wild heterogeneity of the measurements
collected from the literature does not call for any high confidence in the results.
The reported color evolution could have been strongly affected by
contamination to the $V$-band by the extremely strong H$\alpha$ emission
displayed by V1493 Aql. Another word of caution about the Bonifacio et al.
(2000) results is the excessively blue color they report for $t_2$,
$B-V$=0.32. After correcting it for the $E_{B-V}$=0.6 derived by Venturini et al.
(2004), the intrinsic color would be $(B-V)_\circ$=$-$0.3, equal to that of
the hottest known stars ($(B-V)_\circ$=$-$0.32 for O5 stars, Fitzgerald
1970), which contrasts with the cool temperature of the
pseudo-photosphere required by the very low ionization condition of the
emission line spectrum. The discrepancy would be even greater if the
$E_{B-V}$=1.5 reddening estimated by Arkhipova et al. (2002) would be
adopted instead. The $(B-V)_\circ$=$-$0.3 color at $t_2$ also
sharply contrasts with the average $(B-V)_\circ$=$-$0.02 derived by van den
Bergh and Younger (1987) and also perfectly matched by NCyg06 (cf. Sect.~4
above). In conclusion, we are highly suspicious of the odd color
evolution reported by Bonifacio et al. (2000) for V1493 Aql, and consider it
worth further investigation.

\section{Photo-ionization analysis}

A photo-ionization analysis of NCyg06 spectrum for January 20, 2007 ($\Delta
t$=+290$^d$) has been performed with the CLOUDY code, version c06 (Ferland
et al. 1998). This is the spectrum (plotted in Fig.~14) during the nebular
phase with the highest S/N and lower line blending, before the flaring of
the line profile widths seen at later phases (cf. Fig.~12). A detailed
modeling of all collected spectra is far beyond the scope of this paper, and
will be considered elsewhere. Nevertheless, the results we obtained for the
January 20, 2007 spectrum are confirmed by preliminary CLOUDY analysis of
the other spectra, especially concerning chemical abundances.

We assumed spherically expanding ejecta with a gas density declining as
$\rho(r)$$\propto$$r^{-2}$. We explored other density profiles
($\rho(r)$$\propto$$r^n$, with $n$=0,$-$1, and $-$3), but obtained a
poorer match with observations. We did not fix inner and outer radii for
the ionized shell ($r_{in}$ and $r_{out}$), and treated them as free
parameters along with the ratio of the filled to the vacuum volumes in the
ejecta ($\xi$). Only the abundances of chemical elements with observed lines
were allowed to change, and all others were kept fixed to their solar value.
Table~7 lists the observed and computed fluxes for the emission lines used
in the modeling, Table~8 summarizes the modeling results that are
discussed below, and Table~9 provides the corresponding chemical 
mass fractions.

  \begin{figure} 
    \centering 
     \includegraphics[width=8.8cm]{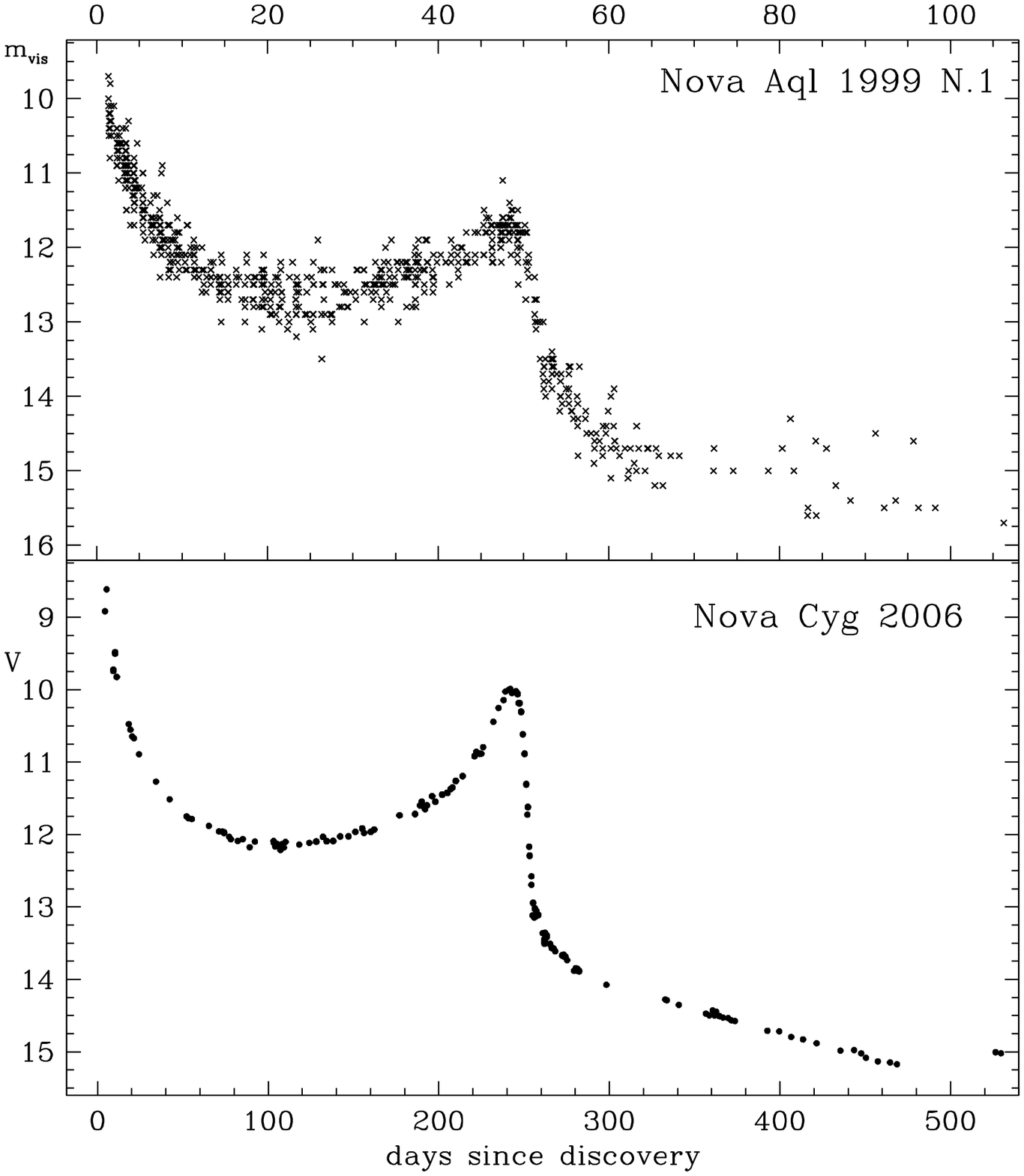}
     \caption{Comparison of Nova Cyg 2006 $V$-band lightcurve with the $m_{\rm vis}$ 
              lightcurve of Nova Aql 1999 N.1 (AAVSO visual estimates). The ordinate 
              scale is the same in both panels. The time scale of Nova Aql 1999 N.1 
              abscissas is 5.0$\times$ faster that that of Nova Cyg 2006.}
     \label{fig15} 
     \end{figure}

  \begin{table}
     \caption{De-reddened emission line ratios relative to H$\beta$ as
              observed on the Jan 20, 2007 spectrum of Nova Cyg 2006 and as
              predicted by the CLOUDY model detailed in Table~8.}
     \centering
     \includegraphics[width=5.2cm]{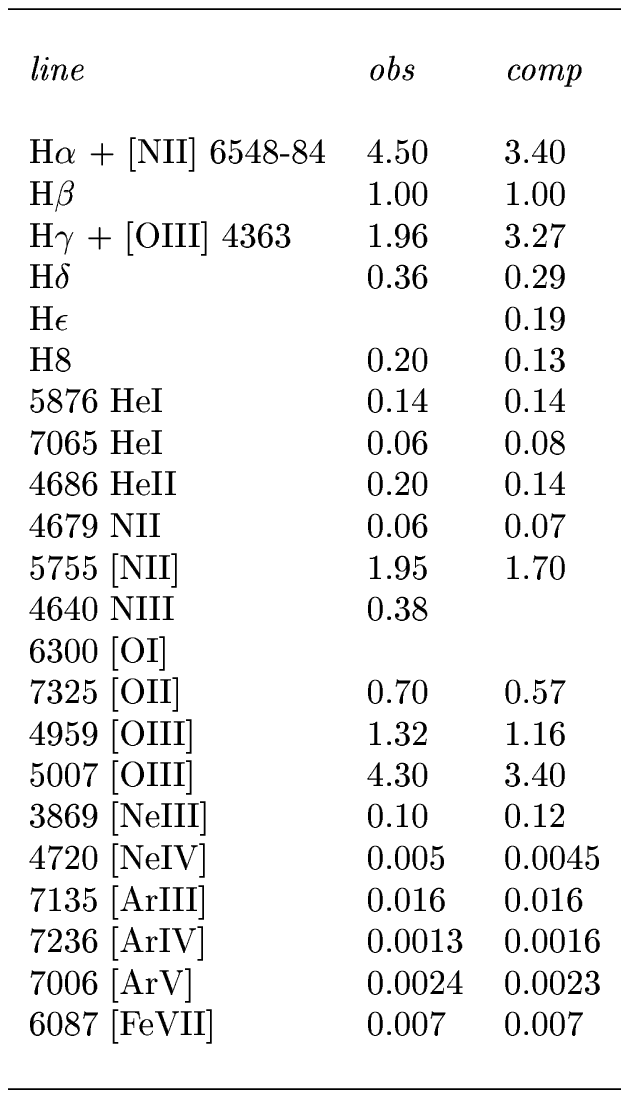}
     \label{tab7}
  \end{table}

  \begin{table}
     \caption{Parameters of the CLOUDY model best-fitting the de-reddened 
              emission line ratio of Table~7. From top to bottom: temperature
              and radius of the central black-body source, inner and outer radii
              of the emitting shell, hydrogen density at inner and outer radii,
              electronic density at the inner and outer radii, ratio of the filled 
              to vacuum volumes, element abundances relative to solar.}
     \centering
     \includegraphics[width=3.8cm]{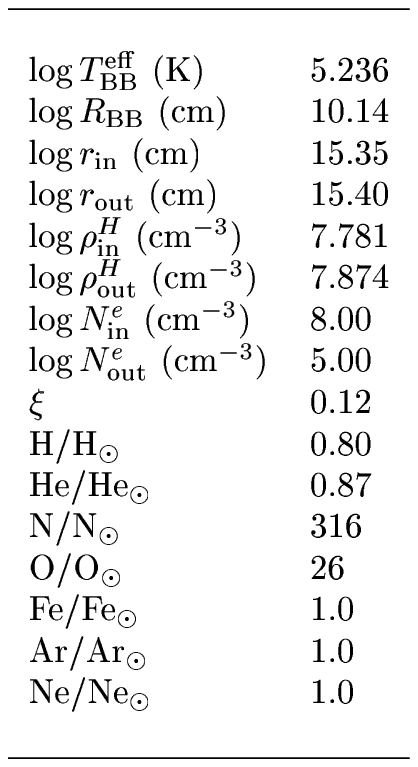}
     \label{tab8}
  \end{table}

  \begin{table}
     \caption{Mass-fraction abundances of measured elements in Nova Cyg 2006 and, 
              for reference, in the Sun.}
     \centering
     \includegraphics[width=3.8cm]{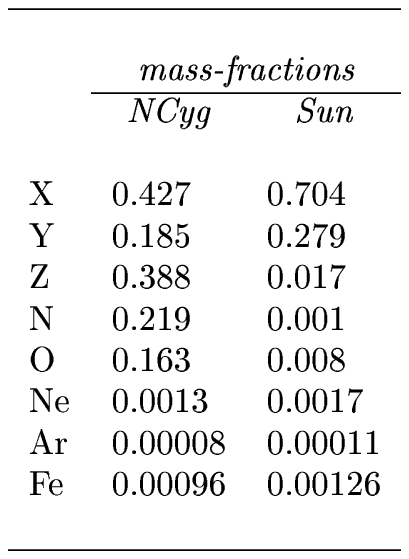}
     \label{tab9}
  \end{table}

\subsection{Internal and external radii}

The shell of ionized gas at $\Delta t$=+290$^d$ is a thin one, extending
from $r_{in}$=149 to $r_{out}$=168~AU. The inner radius is a density
boundary (no interior matter). The outer one is an ionization boundary,
defined by the distance at which the electronic temperature drops to
negligible values (all ionizing photons are absorbed within $r_{out}$, and
only neutral material could exist outward).  These radii correspond to
expansion velocities of 890 and 1010 km~sec$^{-1}$. It is worth noticing
that this is within the range of velocities displayed by the {\em p$_1$
principal} systems of the multi-component absorption spectrum, that it was
the first to appear, the longest lived, and the one carrying the largest
mass outflow, as Fig.~6 and Table~4 clearly illustrate. Therefore, the
ejected material causing the {\em p$_1$ principal} absorption systems during
early post-maximum decline appears to be the same, creating most of the
emission line fluxes at $\Delta t$=+290$^d$ during the nebular phase.

\subsection{Mass in the shell}

The hydrogen mass fraction in the ionized shell is $X$=0.427,
and the ratio of the filled to vacuum volumes is $\xi$=0.12. Therefore, 
the total gas mass within the ionized shell at $\Delta t$=+290$^d$ is
\begin{equation}
M_{shell} = \frac{\xi}{X}\int_{r_{in}}^{r_{out}} 4 \pi r^2 \rho (r) dr = 
3 \times 10^{-4}  ~~{\rm M}_\odot
\end{equation}
In Sect.~5.3.3 we estimated from the [OI] lines the amount of neutral gas
in the ejecta during the early days of the decline from maximum when the
ejecta were predominantly neutral, and found $M_{\rm ejected}^{\rm total}$ = 5
$\times 10^{-4}$ M$_\odot$ (Eq.~5). The agreement of the two estimates is
quite reassuring, in particular if it is taken into account that the ionized
shell at $\Delta t$=+290$^d$ is radiation-bounded and neutral material could
exist outward. Therefore $M_{\rm shell}$ = 3 $\times 10^{-4}$ M$_\odot$ is
a lower limit to the total amount (ionized + neutral) of gas existing at
$\Delta t$=+290$^d$. For sake of discussion in the rest of the paper, we
assume for the total amount of ejected material an averaged value between
the two methods:
\begin{equation} 
M_{\rm ejected}^{\rm total} = 4 \times 10^{-4} ~~{\rm M}_\odot 
\end{equation}

\subsection{Central star}

The central ionizing source is found to have a radius
$R$=0.198~R$_\odot$, a temperature $T_{\rm eff}$=172\,000~K, and therefore a
luminosity 3.08$\times 10^{4}$~L$_\odot$, corresponding to $M_{\rm
bol}$=$-$6.45 (assuming $M_{\rm bol}$=4.74 for the Sun, Livingston 2000).

\subsection{Mass of the white dwarf}

At the time of the January 20, 2007 spectrum, NCyg06 was undergoing the
luminosity plateau phase. It characterizes the observed and theoretical
evolution of novae after the initial super-Eddington phase and before the
switching-off of the nuclear reactions and consequent cooling track for the
white dwarf (cf. Gehrz et al. 1998). During the luminosity plateau phase, the
white dwarf burning residual hydrogen at the surface follows the Paczy\'{n}ski
(1971) core-mass-luminosity relation $L_{\rm cml} = 6 \times 10^4
(M_{\rm WD}/M_\odot - 0.522)$ L$_\odot$. Inserting the above estimated
luminosity, the mass of the white dwarf turns out to be
\begin{equation}
M_{\rm WD} = 1.03 ~~{\rm M}_\odot
\end{equation}
The amount of material to be accreted on this white dwarf for the pressure
to reach 10$^{20}$ dyne and trigger a TNR event is 4$\times
10^{-4}$~M$_\odot$ (cf. Starrfield 1989). In comparison with Eq.(9), the
outburst therefore seems to have ejected most of material
accreted by the white dwarf before the eruption set in.

\subsection{Chemical abundances}                                     

The chemical abundances in the ionized shell are given in Table~8 (relative
to solar) and Table~9 (as mass fractions). They clearly reflect the
non-equilibrium CNO-cycle burning of hydrogen, with massive overabundance of
oxygen and nitrogen, among the largest ever measured in novae. Carbon should
be enhanced too, but there were no detectable carbon emission lines in the
January 20, 2007 spectrum of NCyg06 that could be used to measure its
abundance.

Iron is not enhanced. Because it is not produced by the nuclear reactions,
it is expected to reflect the metallicity of the donor star, which should be
around solar given the likely association of NCyg06 with the galactic disk at
$R$=11.4~kpc galacto-centric distance.

The large overabundances of Ne observed in the spectra of some of the
fastest novae are ascribed to mixing into the accreted envelope of material
from the underlying white dwarf (e.g. Gil-Pons et al. 2003). In massive
progenitors, non-degenerate carbon ignition leads to the formation of a
degenerate core mainly made of oxygen and neon. The minimum mass on the zero 
age main sequence leading to extensive carbon-burning is $M$$\sim$9.3~M$_\odot$ and
the resulting white dwarf will have a mass of $M_{\rm WD}$$\geq$1.1~M$_\odot$.
Neon is not enhanced in NCyg06, in agreement with the above estimate of the
white dwarf mass, which is below the limit for the ONe type.

\section{Energy budget}

The total energy budget of the outburst is the sum of the energy radiated
from the central star and the binding and kinetic energy of the ejecta.

The energy radiated by the central star was computed by integrating the
bolometric luminosity over time. For a rough order of magnitude
estimate, we assumed that the $M_{\rm bol}$=$-$8.0 value at maximum declined
linearly with time to $M_{\rm bol}$=$-$6.45 on $\Delta t$=+290$^d$,
remaining there until the end of our observing campaign on $\Delta
t$=+530$^d$.  The result is
\begin{equation}
E_{rad} = \int_{0}^{530} M_{bol}(t) dt = 1\times 10^{46}~~~{\rm ergs}
\end{equation}
The actual time evolution of the bolometric luminosity is hard to define.
Before NCyg06 settled to the constant luminosity phase, represented by
$M_{\rm bol}$=$-$6.45 on $\Delta t$=+290$^d$, it went through a short-lived,
super-Eddington phase characterized by $M_{\rm bol}$=$-$8.0 at peak $V$-band
brightness. This is a common feature of observed novae and theoretical
models (e.g. Starrfield 1989). Contrary to most novae, NCyg06 went through a
second maximum, and this could have been reflected in an increase in $M_{\rm
bol}$ above its plateau level. Thus a linear trend in luminosity evolution
is adequate to derive an order of magnitude for the radiated energy, which
is moreover a marginal term in the overall energy budget.

The kinetic energy of the ejected shell is 
\begin{equation}
E_{kin} = \frac{\xi}{0.43}\int_{r_{in}}^{r_{out}} 2 \pi r^2 \rho (r) 
\left( \frac {r}{t} \right)^2 dr =  2.2\times 10^{47}~~{\rm ergs}
\end{equation}
where $\rho (r) = \rho$($r_{\rm in}$)($r$/$r_{\rm in}$)$^{-2}$ and the
velocity of the ejecta is taken to be $v(r) = r/t$ which is consistent with
an ejection over a short period of time around $t$=0.

The outburst also had to provide the energy required to gravitationally
unbound the ejected material. The binding energy depends from the mass of
the white dwarf (its potential well). The binding energy of the above
estimated $M_{\rm ejected}^{\rm total}$=$4 \times 10^{-4}$~M$_\odot$ from a
1.03~M$_\odot$ white dwarf is
\begin{equation}
E_{bin} = 2\times 10^{47}~~~{\rm ergs}
\end{equation}
which is quite similar to the kinetic energy. As is typical of novae, most of the
outburst energy goes into the mechanical work of mass ejection, and only a
few percent makes up the radiated fraction.
 
The total energy released by the outburst is therefore
\begin{equation}
E^{tot} = E_{bin} + E_{rad} + E_{kin} = 4.3 \times 10^{47} ~~{\rm ergs}
\end{equation}
which corresponds to the hydrogen burning of 4.6$\times 10^{-5}$~M$_\odot$ of accreted 
material of solar composition, which is about 10\% of the amount of mass accreted by 
the white dwarf prior to the outburst.


\end{document}